\documentclass{IEEEtran}
\usepackage{cite}
\usepackage{amsmath,amssymb,amsfonts}
\usepackage{graphicx}
\usepackage{textcomp,nicefrac}
\usepackage{xcolor}
\usepackage{lipsum}
\usepackage[hidelinks]{hyperref}

\usepackage{booktabs}
\usepackage{multirow}
\usepackage{multicol}
\usepackage{bigstrut}
\usepackage{MnSymbol}
\usepackage{subcaption}
\def\BibTeX{{\rm B\kern-.05em{\sc i\kern-.025em b}\kern-.08em
\T\kern-.1667em\lower.7ex\hbox{E}\kern-.125emX}}

\begin{document}
\bstctlcite{IEEEexample:BSTcontrol}
\title{Single Event Effects Assessment of UltraScale+ MPSoC Systems under Atmospheric Radiation}

\author{Dimitris Agiakatsikas,  Nikos Foutris, Aitzan Sari, Vasileios Vlagkoulis, Ioanna Souvatzoglou, Mihalis Psarakis, Ruiqi Ye, John Goodacre, Mikel Luj\'an, Maria Kastriotou, Carlo Cazzaniga, and Chris Frost 
	\thanks{Manuscript received XX/XX/XX; revised XX/XX/XX; accepted XX/XX/XX. Date of publication XX/XX/XX.}
 \thanks{Experiments at the ISIS Neutron and Muon Source were supported by the beamtime allocation RB2000230 from the Science and Technology Facilities Council. This  work  has also been partially supported by the University of Piraeus Research Center and the EU Horizon 2020 EuroEXA 754337 grant. (\textit{Corresponding authors:} Dimitris Agiakatsikas; Mihalis Psarakis.)}
	\thanks{Dimitris Agiakatsikas (e-mail: agiakatsikas@gmail.com), Aitzan Sari, Vasileios Vlagkoulis, Ioanna Souvatzoglou, and Mihalis Psarakis (e-mail: mpsarak@unipi.gr) are with the Dept. of Informatics, University of Piraeus, Greece.}
	\thanks{Nikos Foutris, Ruiqi Ye, John Goodacre and Mikel Luj\'an are with the Dept. of Computer Science, The University of Manchester, UK.}
	\thanks{Maria Kastriotou, Carlo Cazzaniga and Chris Frost are with the ISIS Facility, STFC, Rutherford Appleton Laboratory, Didcot OX110 QX, UK.}
 \thanks{\copyright2023 IEEE. Personal use of this material is permitted. Permission from
IEEE must be obtained for all other uses in any current or future media,
including reprinting/republishing this material for advertising or promotional
purposes, creating new collective works for resale or redistribution to servers
or lists or reuse any copyrighted component of this work in other works.
This paper is under review by IEEE Transactions on Reliability.}
}
\maketitle
% Instructions for authors for TNS is given in https://ieeexplore.ieee.org/document/1323242
%The abstract should be a short (about 200 words) description of the nature of the problem or topic being discussed and the scope of its treatment: What is the author's unique approach or important contribution, and what is the solution, finding, conclusion, or principal result or application? (Fifty-word (or less) abstracts should accompany Correspondence items.)

\begin{abstract}
The AMD UltraScale+ XCZU9EG device is a Multi-Processor System-on-Chip (MPSoC) with embedded Programmable Logic (PL) that excels in many Edge (e.g., automotive or avionics) and Cloud (e.g., data centres) terrestrial applications. However, it incorporates a large amount of SRAM cells, making the device vulnerable to Neutron-induced Single Event Upsets (NSEUs) or otherwise soft errors. Semiconductor vendors incorporate soft error mitigation mechanisms to recover memory upsets (i.e., faults) before they propagate to the application output and become an error. But how effective are the MPSoC's mitigation schemes? Can they effectively recover upsets in high altitude or large scale applications under different workloads? This article answers the above research questions through a solid study that entails accelerated neutron radiation testing and dependability analysis. We test the device on a broad range of workloads, like multi-threaded software used for pose estimation and weather prediction or a software/hardware (SW/HW) co-design image classification application running on the AMD Deep Learning Processing Unit (DPU). Assuming a one-node MPSoC system in New York City (NYC) at 40k feet, all tested software applications achieve a Mean Time To Failure (MTTF) greater than 148 months, which shows that upsets are effectively recovered in the processing system of the MPSoC. However, the SW/HW co-design (i.e., DPU) in the same one-node system at 40k feet has an MTTF = 4 months due to the high failure rate of its PL accelerator, which emphasises that some MPSoC workloads may require additional NSEU mitigation schemes. Nevertheless, we show that the MTTF of the DPU can increase to 87 months without any overhead if one disregards the failure rate of tolerable errors since they do not affect the correctness of the classification output.   
\end{abstract}


%The text should be succinct: not longer than 20 double-spaced typewritten pages if possible.

% Avoid the use of trade and company names and proprietary terms whenever possible. --> We do not follow this.

\section{Introduction}
\label{sec:introduction}

\IEEEPARstart{M}ulti-Processor System-on-Chip (MPSoC) devices with embedded Field Programmable Gate Array (FPGA) logic are used in many applications such as avionics, automotive, medical, telecommunication, and data centres due to their software flexibility and computational efficiency~\cite{MPSoC-hotchips}. Although semiconductor vendors provide a rich set of MPSoC models, each having different computing capabilities, their fundamental architecture consists of two subsystems integrated into a single chip; the Processing System (PS) and the Programmable Logic (PL), both tightly connected through on-chip high-speed interfaces. The PS subsystem commonly integrates one or more multiprocessors and all kinds of System on Chip (SoC) peripherals like DDR and DMA memory controllers and high-speed (e.g., SATA, PCIe) and general (e.g., Ethernet, CAN, SPI) connectivity interfaces. The PL part is an FPGA, providing the means to implement application-tailored hardware accelerators to offload PS workloads and improve the performance-to-watt ratio metrics of the system.  

Modern MPSoCs store their configuration in Static Random Access Memory (SRAM) cells, which allows them to be reprogrammed practically unlimited times, implementing new or modifying existing hardware accelerators. This is especially useful in data centres, where the processing system of the MPSoC can be reprogrammed multiple times depending on the needs of an application without wearing out the device. However, MPSoCs pose a unique challenge compared to multiprocessor SoCs that do not integrate an FPGA; they are more vulnerable to radiation effects. %and are more power-consuming. 

Unfortunately, reprogramability comes at the expense of a large amount of SRAM cells to store the state and configuration of the device, making it vulnerable to Single Event Effects (SEEs) \cite{Terrestrial-Radiation-FPGA-ACM21, FMER}. Specifically, highly-energised particles like protons originating from deep space (i.e., cosmic rays) and our Sun (i.e., solar rays) collide with nitrogen and oxygen atoms of our Earth's upper atmosphere, producing secondary particles like neutrons and muons \cite{JESD89B, esccsingle}. In turn, neutrons with an energy $\geq$~10~MeV interact with the atoms of the device's semiconductor material, causing SEEs, especially Neutron-induced Single Event Upsets (NSEUs) or otherwise soft errors. NSEUs can upset (i.e., corrupt) the on-chip SRAM memories of the PS, like the multiprocessor's register file and caches~\cite{YANG2019119}, as well as the application (e.g., flip-flops state) and configuration (e.g., programmable routing resources) memory of the PL \cite{Terrestrial-Radiation-FPGA-ACM21}. NSEUs are not permanent, but their effects can threaten system dependability if not well understood and handled in an MPSoC. The failure modes caused by an NSEU range from \textit{unresponsive} errors, for example, an operating system (OS) or program process crash, to \textit{Single Data Corruption (SDC)} errors \cite{mukherjee2011architecture}. SDCs or otherwise an erroneous program output that goes undetected can have catastrophic consequences in an application. For instance, to put an airliner in an uncontrolled steep dive \cite{NSEU-XAPP1073}. 

To fight against the effects of soft errors in MPSoCs, semiconductor vendors incorporate various NSEU mitigation mechanisms in their devices. This brings the following questions: Can the MPSoC's embedded mechanisms effectively mitigate soft errors under all environmental conditions and workloads? For example, what is the Mean Time To Failure (MTTF) of an MPSoC application when it operates at high altitude where the radiation flux can be 500\texttt{x} greater than at sea level~\cite{NSEU-XAPP1073} or when a system uses multiple MPSoCs in large-scale infrastructures like data centres? Are there any types of MPSoC applications that can achieve high MTTF despite an increased rate of memory upsets? This article aims to answer the above research questions with a solid methodology that entails accelerated neutron radiation testing and dependability analysis.

Accelerated radiation testing is the standard and accurate way to trigger neutron-induced SEEs in Integrated Circuits (ICs) to measure their cross-section or otherwise their vulnerability to radiation-induced events. In this work, we exposed a popular MPSoC, the AMD UltraScale+ XCZU9EG, to an accelerated radiation source closely resembling Earth's neutron spectrum for high energies (e.g., $\geq 10MeV$)  to characterise its sensitivity to SEEs. The measured data were then projected and scaled to the expected neutron flux of a target environment to estimate dependability metrics like MTTF of the IC under different workloads and configurations. 
The radiation experiments were performed at ChipIr~\cite{chipir1}, an ISIS neutron and muon facility instrument at the Rutherford Appleton Laboratory, UK.  

Compared to previous works that have performed accelerated radiation testing on the XCZU9EG~\cite{Xilinx_XCZU9EG_Neutrons_protons_tests_2017, ug116_2022, ChipIR_2018_XCZU9EG_tests, anderson_redw_2018}, we make the following contributions:
\begin{itemize}
    \item The MPSoC is tested on a broader range of workloads that exercise the device more exhaustively to reveal more accurate FIT rates than those reported in the literature. We evaluate the cross sections of single-threaded software-only (SW-only) benchmarks that run bare to the metal and complex SW-only Linux-based multi-threaded applications used in weather prediction and pose estimation algorithms. Finally, we irradiated a software-hardware (SW/HW) co-design application, specifically the AMD Deep-learning Processing Unit (DPU) running image classification.
    \item The measured cross-sections of each application are examined under the lens of MTTF and average upset rate, assuming a one-node MPSoC system operating at sea level (e.g., automotive) or 40k feet (airliner's avionics) as well as a 1000-node MPSoC system (e.g., data centre). This helps us understand how well the embedded soft error mitigation mechanisms of the XCZU9EG cope with radiation effects in various terrestrial environments, workloads, and device deployments.
    \item We evaluate the MTTF of the MPSoC for workloads that are inherently resilient to errors.
    \item A fine-grain cross-section characterisation of the PS's Cortex-A53 processor caches and PL memories is provided. For example, we report cross-sections of L1 data and L1 instruction caches, while previous works provide only their average cross-section. 
\end{itemize}

Our results show that the MPSoC will experience, on average, one upset in its PS or PL memories every 24k and 904 months when operating as a one-node MPSoC system in New York City (NYC) at sea level. However, the average upset rates of the PL and PS memories increase to 1.81 and 48 months per upset, respectively, when the same system operates at 40k altitude and doubles in the 1000-node MPSoC system at sea level.
Notably, most of the PS upsets are successfully recovered by the soft error mitigation mechanisms of the MPSoC, ensuring a reliable execution of the SW-only workloads without many SDCs or processor crashes. For instance, all tested SW-only applications  achieve MTTF~$\geq$~148~months, assuming the one-node MPSoC system at 40k feet. However, the SW/HW co-design in the same system has MTTF~=~4~months due to the high FIT rate of its PL DPU accelerator. This points out that some SW/HW MPSoC applications operating at high altitudes or on a large scale may need additional soft error mitigation techniques (e.g., hardware redundancy) to improve reliability. Nevertheless, we show that the MTTF of the DPU application can be improved by 22\texttt{X} if one omits the FIT rate of tolerable output errors since these do not play any role in the correctness of the final classification result.

The rest of the paper is organised as follows. Section~\ref{sec: Background and related work} provides background on the effects of neutron radiation in ICs, and related work of previous accelerated radiation tests of the AMD UltraScale+ MPSoC. Section~\ref{sec:Experiments-Overview} outlines the experimental methodology, radiation test facility, and target boards we used during the experiments. Sections~\ref{sec:BasicTests}~and~\ref{sec:Complex Tests} detail the experimental setup, methodology and results of the MPSoC designs and applications we evaluated under  accelerated neutron radiation testing. Section~\ref{sec:Accessing the reliability of the MPSoC} accesses the reliability of the applications in various environmental conditions and device deployments. Section~\ref{sec:Conclusions} presents concluding remarks.

\section{Background and Related work}
\label{sec: Background and related work}
In this section, we provide the necessary background to understand how atmospheric neutrons can reduce the reliability of MPSoC terrestrial applications.
%and how voltage scaling improves computational efficiency but increases the radiation-induced soft error rate in MPSoCs. 
We also report results from  previous works in atmospheric-like neutron radiation experiments for  AMD 16nm FinFET MPSoCs.

\subsection{AMD 16nm FinFET XCZU9EG MPSoC} 
\label{subsec:AMD XCZU9EG}
The AMD 16nm FinFET XCZU9EG MPSoC is a computing platform that incorporates highly-reconfigurable processing elements to excel in many Edge and Cloud applications. 
As mentioned, the device integrates the following: 1) a Processing System (PS) that incorporates a quad-core Arm Cortex\textsuperscript{TM}-A53 Application Processing Unit (APU) running up to 1.5GHz, 2) a dual-core Arm Cortex\textsuperscript{TM}-R5F real-time processor, 3) an Arm Mali\textsuperscript{TM}-400 MP2 graphics processing unit and 4) Kintex-7 Programmable Logic (PL). The PS is the heart of the MPSoC, including on-chip memory, external memory interfaces, and a rich set of peripheral connectivity interfaces. The XCZU9EG features NSEU mitigation schemes in 1) the PS, e.g., parity check and Single Error Correction Double Error Detection (SECDED) in the APU caches and the on-chip memory (OCM), and 2) the PL configuration and application memories via SECDED mechanisms and layout interleaving schemes to mitigate the effects of multi-bit upsets (MBUs).  

\subsection{Cross-section and failure rate of digital integrated circuits}
\label{subsec: Cross-section and Failure Rate of Semiconductor Devices}
Many Integrated Circuits (ICs) operating in large-scale or high-reliability systems are tested with accelerated radiation experiments to characterise their static and dynamic cross-section under various types of highly-energised particles, like alpha or neutrons.   
The static cross-section quantifies the probability of a Single Event Effect (SEE) occurring when highly-energised particles like neutrons collide with the nucleus of semiconductor material. Mathematically stated: 
\begin{equation}
    \label{eq: Static cross-section}
    \text{Cross-section} = \frac{\text{Number of Events}}{\text{Particle Fluence}} = \frac{\#\text{events}}{\Phi},
\end{equation}
where fluence (represented by the upper-case symbol $\Phi$) defines the number of particles incident on a surface in a given period divided by the area of the surface. 
The larger the static cross-section, the more likely a particle will react with the semiconductor material of the device and the more vulnerable it will be to radiation-induced events like memory upsets.

Once one characterises the static cross-section of a target device, say the NSEU cross-section, it is easy to calculate the expected SER of a device for a given particle flux. For example, the average neutron particle flux in NYC at sea level is approximately 13 neutrons per cm\textsuperscript{2} per hour \cite{JESD89B}, which yields the following Failures In Time (FIT) rate:
\begin{equation}
    \text{FIT} = \text{Static cross-section} \times \frac{13 \text{ neutrons}}{\text{cm}^2 \times \text{hour}} \times 10^9  \text{ hours},
\end{equation}
that is, the average number of failures (e.g., number of memory upsets) that occur within one billion hours of operation \cite{JESD89B}. 

However, not all radiation effects cause an observable error or a system crash in an MPSoC application~\cite{lesea2014soft}. For example, a configuration upset in an unused Look Up Table (LUT) of the PL will probably not affect the operation of a hardware accelerator~\cite{FPGA14}. A memory upset in a register of the APU that is not read but re-written by a new value during the execution of an application will likely not introduce an error~\cite{mukherjee2011architecture}. In a nutshell, not all radiation-induced events (e.g., memory upsets) lead to an application error (e.g., SDC). The dynamic cross-section captures the likelihood of application errors (i.e., only faults that resulted in an output error) for a given particle fluence. It can be calculated with \eqref{eq: Static cross-section} by substituting the number of events with the number of application errors.   

Practitioners that want to assess their reliability in terms of Mean Time To Upset (MTTU) or Mean Time To Failure (MTTF), in other words, the average rate at which memory upsets or application errors occur, can apply the following simple conversion: MTTU or MTTF~[~hours~]~=~1E9~/~FIT.

\subsection{Neutron-induced failures in MPSoC-based terrestrial applications}
\label{subsec:NSEUs-in-terrestial-apps}

Fortunately, most MPSoC terrestrial applications would not experience failures due to atmospheric neutron radiation. The sensitivity per device to NSEUs is extremely low \cite{Terrestrial-Radiation-FPGA-ACM21}. However, the radiation effects increase dramatically when MPSoCs are used on large-scale applications (e.g., data centres) or when operating in high-altitude (e.g., airliner's avionics). Specifically, the rate of NSEU increases for the following reasons.

\subsubsection*{The number of utilised devices in the application increase} Deploying large-scale data centre applications on hundreds of thousands of MPSoCs, collectively increases the total susceptibility of radiation-induced errors over all utilised devices in the system. In other words, if the FIT rate of one ICs is $X$, the overall FIT rate of a system incorporating $N$ such ICs will be FIT\textsubscript{overall}~=~$ X \times N$. In~\cite{Terrestrial-Radiation-FPGA-ACM21}, the authors estimated that the MTTF due to neutron-induced errors on a hypothetical one-hundred-thousand-node FPGA system in Denver, Colorado, would be 0.5 to 11 days depending on the workload.
Indeed, projections from technology evolution roadmaps indicate that the MTTF of data centre computing systems may reach a few minutes \cite{Exascale-2014-Update}. Given that the demand for FPGAs in cloud and data centre facilities will increase in the upcoming decade, and the likelihood of NSEU-related failures may become a significant problem \cite{FPGA-Acceleration-Datacenters-ACM22}. 

\subsubsection*{The device operates at high altitudes} For example, an avionics system at a flight path above 60~$\deg$ latitude at 40k~feet altitude would experience approximately 500 times larger neutron flux than if the same system was operating in NYC sea level \cite{NSEU-XAPP1073}. As we show in section~\ref{sec:Accessing the reliability of the MPSoC}, the average upset rate (i.e., MTTU) of PL memories in an XCZU9EG MPSoC at NYC sea level is 75 years when using the static cross-sections measured in this work. However, using the same device at 60~$\deg$ latitude and 40k~feet altitude will increase the upset rate of the memories to one upset per 1.8 months.   
As mentioned, not all upsets will lead to an error since practical designs commonly do not utilise 100\% of their resources, and some upsets are logically masked during circuit operation \cite{lesea2014soft, mukherjee2011architecture, FPGA14}. Nevertheless, given the tens of thousands of flights per day, the possibility of an SRAM cell upset impacting the safety of a flight is high if the necessary soft error mitigation schemes on the MPSoC design are not in place.

\subsection{Characterisation of the AMD XCZU9EG MPSoC under acceleated atmosperic-like radiation testing}
\label{subsec: Characterisation of the AMD XCZU9EG MPSoC Under Neutron Radiation}
Previous works have tested the AMD XCZU9EG MPSoC with highly-energised ($\geq$10~MeV) neutron and 64~MeV mono-energetic proton accelerated radiation experiments. A 64~MeV mono-energetic protons source approximates the atmospheric neutrons spectrum well and has a lower beamtime cost than neutron beam~\cite{protons_64MeV_XCZU9EG_Xilinx}. However, highly-energised neutrons model more precisely the atmospheric radiation environment and are generally preferred for characterising the cross-section of ICs.

AMD characterised the XCZU9EG MPSoC under neutron at Los Alamos Neutron Science Center (LANSCE) weapons neutron research facility and mono-energetic-protons at Crocker Nuclear Laboratory \cite{protons_64MeV_XCZU9EG_Xilinx}. 
The PS and PL components of the XCZU9EG were exercised with 
the Xilinx proprietary System Validation Tool (SVT)~\cite{protons_64MeV_XCZU9EG_Xilinx}, which executed hundreds of tests per second, resulting in high test coverage. The authors concluded that the CRAM and BRAM static cross-section per bit of the XCZU9EG was reduced by 20\texttt{X} and 16\texttt{X}, respectively, compared to the AMD Kintex-7 FPGA that uses 28nm TSMC's HKMG process technology. In terms of MBUs, 99.99\% of the events were correctable due to the interleaving layout of the MPSoC. The PS was  very reliable, with an overall 1~FIT calculated by projecting the measured cross-sections during the radiation tests to the neutron flux of NYC at sea level. Interestingly, no unrecoverable event in the PS's SRAM structures was reported. All accelerated radiation tests conducted by AMD are officially reported in their UG116 device reliability user guide \cite{ug116_2022}.

Christian Johanson et al. performed neutron radiation experiments on the XCZU9EG MPSoC at ChipIR~\cite{ChipIR_2018_XCZU9EG_tests}. The authors instantiated the AMD Soft Error Mitigation (SEM) IP~\cite{SEM} to collect and post-analyse reports regarding upsets in the device's configuration memory. The BRAMs were initialised with predefined patterns and compared with a golden reference to detect application memory upsets. 

The most comprehensive accelerated neutron radiation testing results for the XCZU9EG have been reported in \cite{XCZU9EG_Wirthlin_SEL_2018} and \cite{anderson_redw_2018} by the \textit{Configurable Computing Laboratory}
of Brigham Young University (BYU).
Specifically, Jordan D. Anderson et al. conducted neutron radiation experiments at LANSCE facility to characterise the NSEU cross-sections of 1) PL memories (i.e., CRAM and BRAM), 2) baremetal single-threaded and Linux-based multi-threaded benchmarks running on the APU (each core run a Dhrystone benchmark -- see Lnx/Dhr in Table~\ref{table: previous XCZU9EG cross-sections}), and 3) APU  memories (i.e., OCM and caches). Notably, the authors did not identify any SDC or processor hang errors during the tests of the APU benchmarks but stated that more beamtime (i.e., fluence) might have been required to obtain statistically significant results \cite{anderson_redw_2018}.  
David S. Lee et al. from the same group characterised the single-event latch-up (SEL)~\cite{JESD89B} cross-section of the XCZU9EG MPSoC under neutrons at LANSCE. The authors tested a technique to detect and recover SELs by monitoring the PMBUS-interfaced power regulators of the ZCU102 board that hosted the device. SELs were observed on the device's VCCAUX and the core supply VCCINT power rails, which were successfully detected and recovered by power cycling the device \cite{XCZU9EG_Wirthlin_SEL_2018}.

Table~\ref{table: previous XCZU9EG cross-sections} summarises the PS and PL cross-sections of the XCZU9EG MPSoC collected by accelerated atmospheric-like radiation tests. Please note that although the authors in \cite{anderson_redw_2018} did not observe any SDC or crash during the software tests, they calculated the cross sections by assuming a single error. This is why the dynamic cross-sections for AES, MxM, and Lnx/Dhr in Table~\ref{table: previous XCZU9EG cross-sections} are not zero even though no errors were observed. Also, note that \cite{protons_64MeV_XCZU9EG_Xilinx} does not provide a detailed characterisation of the PS, e.g., SDC or cache cross sections, as is done in \cite{anderson_redw_2018} and this work. 

As mentioned, except for the detailed NSEU characterisation of the embedded memories of the PS and PL, this paper also studies
the behaviour of complex SW-only and SW/HW applications under the presence of NSEUs to analyse: 1) the reliability of UltraScale+ MPSoC-based systems at the application level in terrestrial environments, 2) the effectiveness of the soft error mitigation approaches embedded in the UltraScale+ devices, 3) the reliability of emerging error resilient applications, e.g., deep neural network (DNN) inference or pose estimation.

\begin{table}[htbp]
	\centering
	\caption{Summary of accelerated atmospheric-like radiation experiments for the AMD XCZU9EG MPSoC}
 \resizebox{0.49\textwidth}{!}{
\begin{tabular}{r|l|r|rr|cc}
\multicolumn{1}{c|}{\multirow{4}[5]{*}{Ref.}} & \multicolumn{1}{c|}{\multirow{4}[5]{*}{Source}} & \multicolumn{1}{c|}{\multirow{4}[5]{*}{Fluence}} & \multicolumn{4}{c}{Cross-section} \\
\cmidrule{4-7}      &       &       & \multicolumn{2}{c|}{PL} & \multicolumn{2}{c}{PS} \\
\cmidrule{4-7}      &       &       & \multicolumn{2}{c|}{ [cm\textsuperscript{2}/bit]} & \multicolumn{2}{c}{[cm\textsuperscript{2}/device]} \\
      &       &   [n/cm\textsuperscript{2}]     & \multicolumn{1}{l}{CRAM} & \multicolumn{1}{l|}{BRAM} & \multicolumn{2}{c}{} \\
\midrule
\cite{Xilinx_XCZU9EG_Neutrons_protons_tests_2017}     & p (64 MeV)  & 1.00E+11 & 3.30E-16 & 1.10E-15 & \multicolumn{2}{c}{6.60E-11} \\
\cite{Xilinx_XCZU9EG_Neutrons_protons_tests_2017}     & n ($\ge$10MeV) & 1.00E+11 & 3.40E-16 & 1.10E-15 & \multicolumn{2}{c}{5.40E-11} \\
\cite{ug116_2022}     & n ($\ge$10MeV) & \multicolumn{1}{c|}{-} & 2.67E-16 & 8.82E-16 & \multicolumn{2}{c}{-} \\
\cite{ChipIR_2018_XCZU9EG_tests}     & n ($\ge$10MeV) & 1.00E+10 & 1.10E-16 & 4.10E-16 & \multicolumn{2}{c}{-} \\
\cite{anderson_redw_2018}     & n ($\ge$10MeV) & 3.00E+11\textsuperscript{$\ostar$} & 2.52E-16 & 3.02E-15 & \multicolumn{2}{c}{See * and \textsuperscript{$\star$}} \\
\multicolumn{6}{l}
{\textsuperscript{$\ostar$}Only for CRAM. For the fluence of BRAM and PS-related  tests, see  \cite{anderson_redw_2018}} \\
\multicolumn{6}{l}{*Cross-section [cm\textsuperscript{2}]: AES:7.66E-11, MxM: 2.70E-11, Lnx/Dhr: 3.95E-12} \\
\multicolumn{6}{l}{\textsuperscript{$\star$}Cross-section [cm\textsuperscript{2}/bit]: OCM:1.47E-16, Caches: 1.5E-15} \\
\end{tabular}%
}
	\label{table: previous XCZU9EG cross-sections}
 %\vspace{-0.4cm}
\end{table}%

\section{Experiments Overview}
\label{sec:Experiments-Overview}
%\subsection{Target-device}
%\label{subsec:Target-device}
%We have performed radiation experiments on the AMD Zynq UltraScale+ MPSoC device (part number XCZU9EG-2FFVB1156E). 

%The experiments have been conducted at \textit{ChipIr}, which is an ISIS neutron and muon facility instrument at the Rutherford Appleton Laboratory, UK. 
\subsection{Experimental Methodology Overview}
\label{subsec:Experimental-methodology-overview} 

It is challenging to perform accelerated radiation testing on a complex computing platform like the XCZU9EG MPSoC as it contains multiple components, each affecting the application differently. 
To overcome the mentioned challenge, we executed a bottom-up experimental methodology. Initially, we tested the PL and PS parts of the device separately and then gradually moved to experiments that tested the PS and PL parts in cooperation. Specifically, we first conducted some basic tests to measure the baseline NSEU and Single Event Functional Interrupt (SEFI)~\cite{JESD89B} cross-sections of all PL memories and to evaluate the SDC and crash (i.e., processor hung) cross-section of SW-only single-threaded baremetal benchmarks. After the basic tests, we moved to access higher-complexity applications. In detail, we evaluated the SDC and crash cross-sections of several multi-threaded SW-only High-Performance Computing (HPC) applications and one popular software/hardware (SW/HW) co-design for DNN acceleration.

In summary, we performed accelerated neutron radiation testing on the following applications.
\begin{itemize}
	\item Basic tests:
	\begin{itemize}
		\item A HW-only PL synthetic benchmark that utilises 100\% of the device's PL resources \cite{Vlagkoulis-TNS21-z7000-HeavyIons}. %This benchmark is tested with a nominal voltage supply.
		\item Several SW-only single-threaded baremetal benchmarks, each one having a different computational and memory footprint.
	\end{itemize}
	\iffalse
    \item Undervolting tests:
	\begin{itemize}
		\item The same HW-only PL synthetic benchmark \cite{Vlagkoulis-TNS21-z7000-HeavyIons} that was used for the basic tests but with the VINT supply voltage of the PL fabric undervolted. Please note that we have performed undervolting tests only on the PL part of the device. The PS part has not been tested with an undervolted supply voltage.
	\end{itemize}
    \fi
	\item Complex tests:
	\begin{itemize}
			\item Two complex SW-only multi-threaded applications running under Linux OS. Specifically:
		\begin{itemize}
			\item LFRiC, which is a compute-intensive kernel for weather and climate prediction \cite{LFRiC}.
			\item Semi-direct Monocular Visual Odometry (SVO), which is used in automotive and robotic systems for pose estimation \cite{SVO}. 
		\end{itemize}
		\item One SW/HW multi-threaded co-design application running under Linux OS. Specifically, the AMD Vitis DPU \cite{vitis-1.3.1}, which is a popular Convolution Neural Network (CNN) accelerator. %executing the \texttt{resnet50} image classification model.
	\end{itemize}
\end{itemize}

\subsection{Radiation test facility} 
\label{subsec:Radiation-test-facility} 
We performed the radiation tests at ChipIr at the Rutherford Appleton Laboratory in Oxfordshire, UK. ChipIR is designed to deliver a neutron spectrum as similar as possible to the atmospheric one to test radiation effects on electronic components and devices \cite{chipir1, chipir2}. The ISIS accelerator provides a proton beam of 800~MeV at 40~$\mu$A at a frequency of 10~Hz, impinging on the tungsten target of its target~station~2, where ChipIr is located.
The spallation neutrons produced illuminate a secondary scatterer, which optimises the atmospheric-like neutron spectrum arriving at ChipIr with an acceleration factor of up to 10\textsuperscript{9} for ground-level applications. With a frequency of 10~Hz, the beam pulses consist of two 70~ns wide bunches separated by 360~ns. The beam fluence at the position of the target device was continuously monitored by a silicon diode, while the average flux of neutrons above 10~MeV during the experimental campaign was 5.6E+6~neutrons/cm\textsuperscript{2}/seconds. The beam size was set through the two sets of the ChipIr jaws to 7cm~x~7cm. Irradiation was performed at room temperature.
\figurename~\ref{fig:beam} depicts the target boards we irradiated at ChipIr.

The cross-section calculations in this work assume a Poisson distribution of the NSEUs, a confidence level of 95\%, and 10\% uncertainty on the measured fluence.

\subsection{Target boards}
\label{subsec:Target-board}  
We conducted the radiation experiments on two AMD ZCU102 evaluation boards (revision 1.1), each hosting the XCZU9EG chip. 
One board was modified to disconnect a few onboard switching voltage regulators and power the board with an external multichannel Power Supply Unit (PSU). We modified the board to protect it from Single Evert Latch-ups (SELs) that cause radiation-induced high-current events.
The second board was used \textit{out-of-the-box} for the complex tests. In other words, it was not modified. 

\subsubsection*{Modified ZCU102 board} 
\begin{figure}[t]
	\centerline{\includegraphics[width=3.3in]{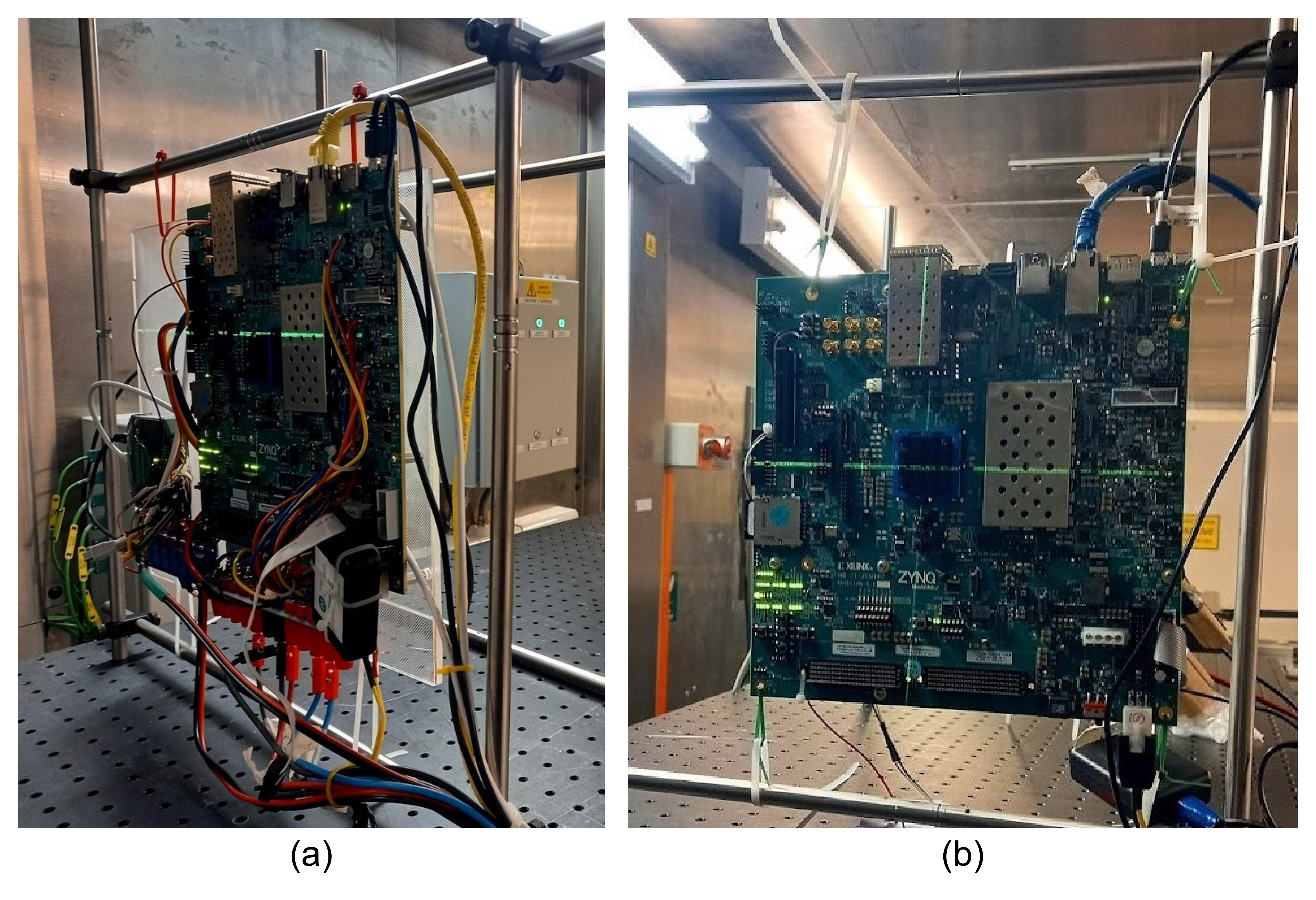}}
	%\vspace*{-2mm}
	\caption{Neutron beam experiment at the ChipIr facility of RAL, UK. \figurename~\ref{fig:beam}(a) shows the modified ZCU102 board with its voltage rails (0.85V, 1V2, 1V8 and 3V3) powered by an external multichannel power supply unit. \figurename~\ref{fig:beam}(b) illustrates the \textit{out-of-the-box} ZCU102 board, which uses its onboard voltage regulators.}
	\label{fig:beam}
        %\vspace{-0.4cm}
\end{figure}

Previous neutron radiation experiments on a ZCU102 board (revision -- engineering sample~1) showed that some onboard voltage regulators are vulnerable to high-current events \cite{lee2018single}. 
To protect the board from these anticipated events, we adopted the solution of David S. Lee et al. \cite{lee2018single}. 
Specifically, we 1) removed all onboard voltage regulators for 3.3V (VCC3v3, UTIL\_3V3), 0.85V (VCCBRAM, VCCINT, VCCPSINTFP, VCCPSINTLP), 1.2V (DDR4\_DIMM\_VDDQ) and 1.8V (VCCAUX, VCCOPS) power rails and 2) provided voltage to the mentioned power rails via a multichannel PSU. A Python script running on a PC (see Control-PC in \figurename~\ref{fig:basic and undervolting tests setup}) monitored the current drawn from each PSU channel to power cycle (i.e., turn off and on) the board during high-current events.
\figurename~\ref{fig:beam}(a) shows the ZCU102 board with its voltage rails (0.85V, 1V2, 1V8 and 3V3) powered by an external PSU.

%Justify the second setup
\subsubsection*{Out-of-the-box ZCU102 board} 
During the preparation of the tests, before the radiation experiments, we observed that the modified board often crashed during the boot time of the Linux OS (i.e., for testing the LFRiC, SVO and AMD DPU applications). The crashes were caused by voltage droops due to an instantaneous (fast) increase of the current at the 0.85V and 1.2V power rails when the Linux kernel was performing the initialisation of the PS DDR memory. Our external PSU setup could not sustain a stable 0.85V and 1.2V power supply during these current spikes.
%To overcome the mentioned problem, we run the Linux-based applications on a out-of-the-box board and performed a periodic power cycle (i.e., every 15 min) as a preventive measure to eliminate possible low impedance paths before causing catastrophic damages.
To overcome the mentioned problem, we ran the Linux-based applications (i.e., complex tests) on the \textit{out-of-the-box} board. We used the PMBUS Maxim Integrated PowerTool as suggested by \cite{lee2018single} to detect SELs. Please note that depending on the target IC, a SEL can cause a rapid increase in the current of a power rail that is difficult to detect on time and power of the device before it is damaged. However, as shown in \cite{lee2018single}, the rate at which current increases in the XCZU9EG power rails during an SEL is slow. This gives plenty of time (commonly a few minutes) to detect and recover a high-current event by power cycling the target board. Although detecting and recovering a high-current event is faster with an external PSU, the experience we gained from these experiments indicates that the PMBUS Maxim Integrated PowerTool is a sufficient solution to protect the board. \figurename~\ref{fig:beam}(b) shows the unmodified ZCU102 board we used for the complex tests.

\section{Basic Tests}
\label{sec:BasicTests}

This section presents the experimental methodology and results of all basic tests. The objectives of these tests are the following: 1) characterise the NSEU and SEFI static cross-sections of all PL memories using synthetic HW benchmarks and 2) evaluate the dynamic SDC and crash cross-sections of several SW-only single-threaded baremetal applications running on the APU.  

\subsection*{Experimental setup and overview for all basic tests:} 
\label{subsec:Test setup for the basic tests}
\begin{figure}[t]
	\centerline{\includegraphics[width=3.5in]{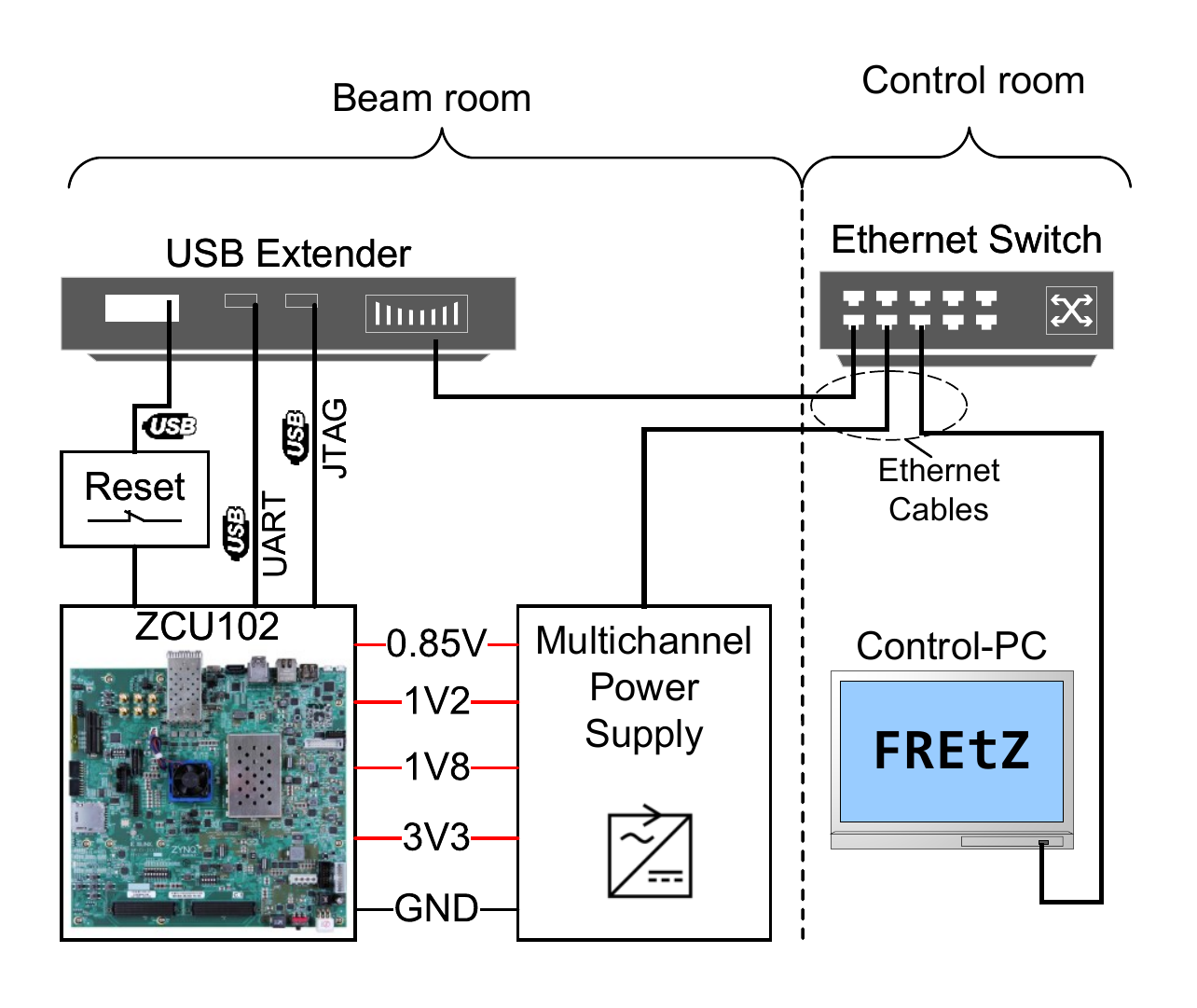}}
	%\vspace*{-6mm}
	\caption{Experimental setup to collect results for the basic, i.e., NSEU and SEFI static cross-section of all PL memories, and SDC dynamic cross-section of several single-threaded baremetal benchmarks running on the APU.}
	\label{fig:basic and undervolting tests setup}
        %\vspace{-0.4cm}        
\end{figure}

\figurename~\ref{fig:basic and undervolting tests setup} presents the setup for the basic tests, which are conducted on the modified ZCU102 board (see section~\ref{subsec:Target-board}). 
Specifically, a computer, namely the Control-PC, is located in the control room and orchestrates the tests by performing the following tasks:
\begin{itemize}
        \item Configures, controls and monitors the execution of benchmarks on the target board.
        \item Resets the board during benchmark timeouts (i.e., radiation-induced events that make the device unresponsive) by electrically shorting the board's \texttt{SRTS\_B} and \texttt{POR\_B} reset buttons via a USB-controlled relay.
	   \item Monitors an Ethernet-interfaced multichannel PSU to power cycle the board during, if any, high-current events.
\end{itemize}
Note that all USB connections are transferred from the beam room to the control room via an Ethernet-based USB extender.

\subsection{HW-only PL synthetic benchmark tests} % 
\label{subsec:HW-only PL synthetic benchmark tests}

\subsubsection*{Benchmark details}
\label{subsubsec:Benchmark details}  
We performed the PL tests on a highly utilised and densely routed design, which instantiates all slice, Block-RAM (BRAM), and Digital Signal Processor (DSP) primitives of the XCZU9EG device. The design has the following characteristics:
\begin{itemize}
 \item All PL slices are combined into multiple long register chain structures. In detail, the LUTs of SLICEL and SLICEM tiles are configured as route-through and 32-bit Shift Register LUT (SRL), respectively. The LUT outputs of all PL slices are connected with their corresponding slice Flip-Flops (FFs) to form long register chains. Each SRL in the device is initialised with predefined bit patterns. 
 \item All BRAMs are cascaded through their dedicated data bus horizontally (i.e., raw) or vertically (i.e., column) and initialised with address-related bit patterns.
 \item Clock and clock-enable signals of all BRAM are set to `0' (i.e., disabled) to reduce the likelihood of BRAM upsets caused by Single Event Transients (SETs) on the clock tree and BRAM data bus signals of the device. We aim to reduce transient upsets since we focus on characterising the NSEU and SEFI cross-section of the device.
 %In detail, the upsets due to transients in the clock tree or the BRAM data buses are unlikely to happen be %Most upsets in BRAMs or LUTRAMs are expected to occur due to SEUs.
 \item All DSP primitives are connected in cascade mode and configured to implement Multiply and ACcumulate (MAC) operations.
\end{itemize}
Detailed information for the tested synthetic benchmark can be found in our previous work \cite{Vlagkoulis-TNS21-z7000-HeavyIons}, where we used the same benchmark to characterise the PL memories of an AMD Zynq-7000 device under heavy ions.

\subsubsection*{Testing procedure}
\label{subsubsec:Testing procedure}   
The Control-PC downloads via JTAG the bitstream of the PL synthetic benchmark into the XCZU9EG device. In turn, it performs readback capture via JTAG \cite{XAPP1230} for 50 consecutive times, each time logging the state of all CRAM and Application RAM (ARAM) (e.g., FFs and BRAM contents) bits of the device in a readback file. This test procedure cycle (i.e., one device configuration and 50 readbacks) is continuously performed until the end of the test. In case of an unrecoverable error, the Control-PC performs the following tasks: 1) power cycles the ZCU102 board via the Ethernet-controlled PSU, 2) reconfigures the device and 3) continues readback capture from where it was left before the radiation-induced event occurred. All events that make the XCZU9EG device unresponsive are classified as unrecoverable. For example, a radiation-induced upset in the JTAG circuitry of the target device may result in a connection loss and make the device unresponsive to all JTAG queries made by the Control-PC. 
 
We should make two notes for the testing procedure of the PL synthetic benchmark:
\begin{itemize}
	\item Accumulated upsets are cleared in the device on average every 1400 seconds, i.e., by downloading the bitstream into the device after 50 continuous readbacks, which last \emph{50~readbacks~$\times$~28 seconds per readback = 1400 seconds}.
	\item All JTAG transactions with the target device are performed by our open-source FREtZ tool \cite{eslab:fretz, sari2019open-fretz}. FREtZ provides a rich set of high-level Python APIs and application examples to readback, verify and manipulate the bitstream and the device state of all AMD 7-series and UltraScale/UltraScale+ MPSoC/FPGAs. Specifically, FREtZ increases the productivity of performing fault-injection and radiation experiments by hiding low-level Vivado TCL/JTAG commands that are executed behind the scenes to access the PS and PL memories of the target device. 
    \item The results of the basic tests are obtained by post-analysis of the collected data (i.e., readback files). Each readback file consists of 1) configuration bits that specify the functionality of the design and device, 2) flip-flop and slice LUTRAM contents, and 3) BRAM contents. Configuration bits are static bits because they do not change during circuit operation, while the flip-flop, LUTRAM, and BRAM contents are dynamic bits, i.e., change during circuit operation, assuming a clock provision.
    AMD Vivado design suite produces a mask file during bitstream generation that FREtZ applies on each readback file to distinguish the static from the dynamic bits when analysing our experimental data and results.
\end{itemize} 

\subsubsection*{Results -- NSEU cross-section of the PL memories}
\label{subsubsec:Cross sections of the PL memories}  
Table~\ref{table:pl-cross-sections} shows the neutron static cross-section and the number of SEFI occurrences of the target device. 
Each PL memory type (CRAM, BRAM and SRL) was exposed to radiation for approximately six hours with 5.6E+6 neutrons/cm\textsuperscript{2}/seconds flux, thus accumulating 1.2E+11 neutrons/cm\textsuperscript{2} fluence on average (see 2\textsuperscript{nd} column of the table). The 1.2E+11 fluence is equivalent to exposing the device to the radiation environment of NYC at sea level for more than 1.3 million hours. 
 In detail, the 3\textsuperscript{rd} column of the table shows the number of upsets for each memory type, while 4\textsuperscript{th} and 5\textsuperscript{th} columns illustrate the cross-section per device and bit, respectively. The CRAM static cross-section that we measured (1.84E-16~cm\textsuperscript{2}/bit) is in the range 1.10~E-16~cm\textsuperscript{2}/bit~--~3.40~E-16~cm\textsuperscript{2}/bit as reported in previous studies and summarised in Table~\ref{table: previous XCZU9EG cross-sections}. The cross-section of BRAM and SRL per cm\textsuperscript{2} per bit is one order of magnitude higher than CRAM, which matches with the findings of AMD~\cite{Xilinx_XCZU9EG_Neutrons_protons_tests_2017} and BYU~\cite{anderson_redw_2018}. 

The last column of Table~\ref{table:pl-cross-sections} shows the number of SEFIs per memory type, which is analysed in the following paragraphs. 

\begin{table}[htbp]
	\centering
	\caption{NSEU cross-section of the PL memories}
	\begin{tabular}{l|r|r|r|r|c}
		\multicolumn{1}{c|}{\multirow{4}[5]{*}{\textbf{Type}}} & \multicolumn{1}{c|}{\multirow{2}[3]{*}{\textbf{Fluence}}} & \multicolumn{3}{c|}{\textbf{NSEU}} & \textbf{SEFIs} \bigstrut[b]\\
		\cline{3-6}          &       & \multicolumn{1}{c|}{\multirow{3}[4]{*}{\textbf{Upsets}}} & \multicolumn{2}{c|}{\textbf{Cross section}} & \multirow{3}[4]{*}{\textbf{\#}} \bigstrut\\
		\cline{4-5}          & \multicolumn{1}{c|}{\multirow{2}[2]{*}{\textbf{[$n/cm^2$]}}} &       & \multicolumn{1}{c|}{\textbf{Device}} & \multicolumn{1}{c|}{\textbf{Bit}} &  \bigstrut[t]\\
		&       &       & \multicolumn{1}{c|}{\textbf{[$cm^2$]}} & \multicolumn{1}{c|}{\textbf{[$cm^2/bit$]}} &  \bigstrut[b]\\
		\hline
		CRAM  & 1.20E+11 & 2,417 & 2.01E-08 & 1.84E-16 & 0 \bigstrut[t]\\
		BRAM  & 1.20E+11 & 10,118 & 8.42E-08 & 1.21E-15 & 1 \\
		SRL   & 1.20E+11 & 1,462  & 1.22E-08 & 1.32E-15 & 1 \\
	\end{tabular}%
	\label{table:pl-cross-sections}
        %\vspace{-0.4cm}
\end{table}%

\subsubsection*{Results -- SBU, MBU and MCU events in the PL memories}

We adopted the statistical analysis approach of \cite{wirthlin2014method} to distinguish NSEUs that caused Single-Bit Upsets (SBUs), Multi-Bit Upsets (MBUs) and Multi-Cell Upsets (MCUs). JEDEC refers to MBUs as multiple upsets occurring in one configuration frame and MCUs expanding in one or more (usually neighbouring) configuration frames \cite{JESD89B}. In general, recovering MBUs with classic Error Correction Code (ECC) based CRAM scrubbing\cite{scrubbing-IEEETranRel} is challenging because each configuration frame of the XCZU9EG embeds ECC information that can only support the correction of an SBU. However, ECC scrubbing can successfully correct MCUs (i.e., multiple SBUs in different configuration frames).  

Table~\ref{table:pl-mcu} presents the percentage of NSEUs that caused an SBU or an MCU, as well as their shapes (i.e., upset patterns). The x-axis of the shapes represents consecutive frames (i.e., frames with consecutive logical addresses), while the y-axis represents consecutive bits in a frame. 

Our results show that approximately 96\% of NSEUs resulted in SBUs and the remaining 4\% in MCUs. The MCUs appear in five shapes as shown in Table~\ref{table:pl-mcu} and extend from 2 to 8 frames, while the bit multiplicity reaches up to 3 bits. 
%However, we observe a trend as VINT decreases; the percentage of total MCUs increases and the MCU events occur in CRAM frames with larger logical address distances. A reduction of VINT results in a reduction of Qcrit in the MPSoC's configuration memory SRAM cells. Therefore, more physically neighbouring cells get corrupted as Qcrit is reduced even that the Linear Energy Transfer (LET) is kept the same during the tests with nominal and undervolted supply voltages. 
Finally, we did not observe any MBU, which can be justified by the memory interleaving features of UltraScale/+ MPSoC devices. This is to say, memory cells belonging to the same logically addressed frame are physically separated, thus mitigating MBUs commonly caused in neighbouring physical cells. The NSEU shape results suggest that SECDED scrubbing is an adequate CRAM error recovery mechanism for XCZU9EG MPSoCs used in terrestrial applications since no MBUs were observed during our accelerated radiation tests.

%Comparing the MCUs between the three VINT supply voltages, 
%Micheal Wirthlim et al. in \cite{Wirthlin2014MCUs} showed that the percentage and size (how far apart MCUs occur) of MCU events increase as Linear Energy Transfer (LET) increases. We did not change the LET of the beam during the undervolting tests. Still, the nominal supply voltage setup required lower LET than the undervolted setups to generate MCUs of larger size, having, therefore, similar effects as changing the LET itself.

%\begin{figure}[t]
%	\centerline{\includegraphics[width=3.5in]{figures/jpeg/mcu.jpg}}
%	\vspace*{-1mm}
%	\caption{PL - MBUs/MCUs in CRAM}
%	\label{fig:mcu}
%\end{figure}

% Table generated by Excel2LaTeX from sheet 'Sheet2'
\begin{table}[htbp]
	\centering
	\caption{NSEU shapes in the CRAM}
	\begin{tabular}{lllllll}
		%\multicolumn{1}{c|}{\multirow{2}[2]{*}{\textbf{VINT}}} & \multicolumn{6}{c}{\textbf{SEU shapes}} \bigstrut[b]\\
		\cline{1-6}    %\multicolumn{1}{c|}{} &
  \multicolumn{1}{c|}{\textbf{SBUs [\%]}} & \multicolumn{5}{c}{\textbf{MCUs [\%]}} \bigstrut[t]\\
		%\multicolumn{1}{c|}{\textbf{[V]}} & 
  \multicolumn{1}{c|}{\includegraphics[width=0.05in]{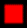}} & \includegraphics[width=0.11in]{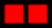} & \includegraphics[width=0.26in]{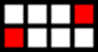} & \includegraphics[width=0.48in]{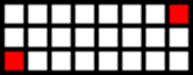}& \includegraphics[width=0.16in]{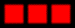} & \includegraphics[width=0.12in]{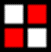} \bigstrut[b]\\
		\hline
		%\multicolumn{1}{r|}{0.85} & 
  \multicolumn{1}{c|}{93.80} & \multicolumn{1}{c}{4.07} & \multicolumn{1}{c}{0.84} & \multicolumn{1}{c}{0.57} & \multicolumn{1}{c}{0.35} & \multicolumn{1}{c}{0.09} \bigstrut[t]\\
		%\multicolumn{1}{l|}{**0.75} & \multicolumn{1}{c|}{93.50} & \multicolumn{1}{c}{4.14} & \multicolumn{1}{c}{0.85} & \multicolumn{1}{c}{0.49} & \multicolumn{1}{c}{0.24} & \multicolumn{1}{c}{--} \\
		%\multicolumn{1}{l|}{**0.60} & \multicolumn{1}{c|}{93.40} & \multicolumn{1}{c}{4.74} & \multicolumn{1}{c}{1.35} & \multicolumn{1}{c}{--} & \multicolumn{1}{c}{--} & \multicolumn{1}{c}{--} \\
		%\multicolumn{7}{l}{*Nominal, **Undervolted} \\
	\end{tabular}%
	\label{table:pl-mcu}
        %\vspace{-0.1cm}
\end{table}%

\iffalse
\begin{table}
\centering
\caption{PL - MCU shapes in CRAM}
\begin{tabular}{l|c|ccccc} 
\hline
\textbf{Test Sessions} & \textbf{SBUs} & \multicolumn{5}{c}{\textbf{MCU shapes}} \\ 
 &  \includegraphics[width=0.07in]{figures/png/sbu.png} & \includegraphics[width=0.14in]{figures/png/mcu1.png} & \includegraphics[width=0.28in]{figures/png/mcu2.png} & \includegraphics[width=0.56in]{figures/png/mcu3.png} & \includegraphics[width=0.21in]{figures/png/mcu4.png} & \includegraphics[width=0.14in]{figures/png/mcu5.png} \\
\hline

\textbf{Nominal (850mV)} & 2121 & 92 & 19 & 13 & 8 & 2 \\
\textbf{Under (750mV)}    & 767  & 34 & 7  & 4  & 2 & - \\
\textbf{Under (600mV)}    & 414  & 21 & 6  & -  & - & - \\

\hline
\end{tabular}
\label{table:pl-mcu2}
\end{table}
\fi
%\subsubsection*{Results -- SEFI cross-section of the PL memories with undervolted VINT voltage supply}
%In contrast to the tests with a nominal supply voltage, we did not observe any SEFI in the CRAM, BRAM and SRL memories during the undervolted tests. However, we believe that more tests (i.e., more fluence) are required to obtain statistically significant results to conclude whether undervolting reduces the likelihood of SEFIs. 
\subsubsection*{Results -- SEFIs in the PL memories}
As shown in Table~\ref{table:pl-cross-sections} we observed two SEFIs during the basic PL tests;

\emph{BRAM SEFI:}~The SEFI exhibited as a multi-bit upset affecting almost all the words of a BRAM. Specifically, all the even-numbered addresses (i.e., 0, 2, \dots, 1022) of a 36Kb BRAM (i.e., 1024 $\times$ (32 data bits + 4 parity bits)) were written with the predefined value of the $1022^{nd}$ word due to the SEFI, while all the odd-numbered addresses (i.e., 1, 3, \dots, 1023) were written with the value of the $1023^{rd}$ word. This BRAM SEFI resulted in 10.5~kb (instead of 36~kb) upsets since many memory addresses were written with their initial value, i.e., the upsets were logically masked. We excluded the upsets caused by the SEFI when calculating the NSEU cross-section of the BRAMs in Table~\ref{table:pl-cross-sections}.

\emph{SRL SEFI:}~We found that a SET on the clock signal in one CLB slice of an SRL caused the SEFI. Specifically, all the 256 SRL bits located in the eight LUTMs of the same slice (each SLICEM consists of eight 32-bit SRLs, and each SRL occupies a 64-bit LUTM in a master/slave arrangement) were corrupted by the SET on their clock signal. Similarly to the BRAM SEFI, the upsets caused by the SRL SEFI are removed from the NSEU cross-section calculations in Table~\ref{table:pl-cross-sections}.  

%In the case of the SEFI in the SRL memory type, the observed upsets could be caused by either an SEU in the LUT bits or a SET in the clock net and reset signals of the corresponding chains (remember that all SRLs are cascaded in long chains). Here, we observed the effects of a clock signal SET in the SRL bits of one CLB slice. Specifically, all the 256 SRL bits located in the eight LUTMs of the same slice (each SLICEM consists of eight 32-bit SRLs, and each SRL occupies a 64-bit LUTM in a master/slave arrangement) were affected by the SET. Similarly to BRAM, the SET-induced upsets were removed from the calculation of the SRL cross-section.

\subsubsection*{Results -- High-current events in the MPSoC}
During the PL tests we observed two high-current events; one occurred at the 1.8V power rail of the MPSoC and one at the 3.3V. 
%In the 0.75V undervolted PL tests, we counted two high-current events at the MPSoC's 1.8V power rail and none when the device was powered with VCCINT=0.60V. 
The high-current events were successfully recovered by power cycling the device. We did not detect any high-current event in the SW-only single-threaded baremetal benchmarks basic tests and all complex tests.

The results of SEFIs and high-current events show that the probability of such phenomena is extremely low; the device may experience, on average, a BRAM SEFI, a SRL SEFI or two high current events after 1.3 million hours, assuming operation in NYC at sea level. In other words, the equivalent time of natural neutron exposure in NYC to achieve the fluence of the accelerated radiation tests.

\subsection{SW-only single-threaded baremetal benchmarks basic tests}
\label{Processing-System-Tests}

\subsubsection*{Benchmarks details}
We executed the following six embedded microprocessor benchmark kernels used in many real-world applications: CRC32, FFT, Qsort, BasicMath, SHA, and MatrixMul. All benchmarks were sourced from the MiBench suite \cite{MiBench}, except MatrixMul, which was developed in-house. MiBench programs were adapted to run on the ARM CPU as baremetal single-threaded applications. 

We selected or modified the benchmark's input data sets to compose programs with different memory footprints, i.e., different data memory segment lengths. In this way, we were able to evaluate the impact per cache level on the SDC and crash rates under different cache utilisation conditions.
The memory footprints of the benchmarks are shown in Table~\ref{table:ps-benchmarks}. The data segment includes global and static variables, while Read Only (RO) data includes constant data. 
%Να θυμάσαι μόνο το εξης: κάποια benchmarks χρησιμοποιούν δεδομένα που είναι global κατά την εκτέλεση τους. Αυτά είναι στο data segment. Κάποια έχουν και constants. Αυτά είναι στο ro. Σίγουρα, όλα χρησιμοποιούν stack και για κάποια benchmarks αρκεί μόνο το stack, οι local μεταβλητές δηλαδή, για την εκτέλεση τους. Αυτά δεν έχουν δεδομένα στο data segment. Θα μπορούσαν όλα τα benchmarks να τα φτιάξουμε να έχουν τα δεδομένα ως public. Αλλά θα έπρεπε να αλλάξουμε όλα τα functions για μερικά bytes που άλλωστε να επηρρεαζαν κ το performance 
One note should be made for the data segment usage of SHA and MatrixMul benchmarks; the SHA and MatrixMul benchmarks have been developed as functions and do not use global and static variables as other benchmarks do. Therefore, all computations for SHA and MatrixMul are performed in local variables. The data segments (stored temporarily in the stack) of the SHA and MatrixMul benchmarks are less than 32~KB and are not reported in Table~\ref{table:ps-benchmarks}. 
%The memory stack is not considered for calculating the data segments of the benchmarks. Therefore, the data segment in the cases of SHA and MatrixMul benchmarks is zero. The behaviour of the caches largely depended on the memory footprint since all benchmarks ran separately on the Arm APU core.

In summary, the benchmarks have the following characteristics:
\begin{itemize}
\item The data segments of the FFT, BasicMath, SHA and MatrixMul fit into the L1 data cache (32~KB) of the APU core. Thus cache conflict misses are unlikely to happen. 
\item The data segment of Qsort does not fit into the L1 data cache (32~KB), but it does fit into the L2 cache (1~MB); this means that during the execution of QSort, several conflict cache misses and thus cache replacements may occur in the L1 cache but not in the L2 cache.
\item The data segment of CRC32 does not fit into the L2 cache; this means that during the execution of CRC32, several replacements in L2 may occur.
\end{itemize}

\subsubsection*{Testing procedure}
The Control-PC shown in \figurename~\ref{fig:basic and undervolting tests setup} communicates with the PS through the PL JTAG interface. The PS stores the benchmark output results in the PS DDR memory, and the Control-PC collects the results through the JTAG interface. In more detail, a JTAG-to-AXI bridge is instantiated into the PL to access the DDR memory through a high-performance AXI port.
%via JTAG commands to AXI transactions. 
The Control-PC uses the same JTAG-to-AXI bridge interface to configure the PS and initiate the execution of the benchmarks.
To guard these auxiliary components (e.g., JTAG-to-AXI bridge) against radiation-induced errors during the tests:  
%Please note that we incorporated  the following two SEU mitigation schemes to reduce the likelihood of radiation-induced errors in the PL circuits of the device: 
1) we instantiated the AMD SEM IP core~\cite{SEM} to correct CRAM upsets, and 2) triplicated all components (including the SEM IP) in the PL with Synopsis Synplify Premier~\cite{SynopsisPremier15}.

\begin{table}
\centering
\caption{CPU benchmarks -- Memory footprints}
\begin{tabular}{lrrr} 
\hline

\textbf{Benchmark}	& \textbf{Code Segment}	& \textbf{RO Data} & \textbf{Data Segment} \\
\hline
FFT        & 2.81 KB	& 0.20 KB	 & 2.09 KB \\
SHA       & 2.14 KB	& 2.32 KB	 & 0.00 KB \\
BasicMath  & 2.74 KB	& 0.10 KB	 & 6.09 KB \\
MatrixMul  & 0.77 KB	& 23.74 KB	 & 0.00 KB \\
Qsort      & 0.25 KB	& 512.00 KB & 156.25 KB \\
CRC32      & 0.57 Kb	& 0.00 KB	 & 2675.56 KB \\
\hline
\end{tabular}
\label{table:ps-benchmarks}
\vspace{0cm}
\end{table}

\subsubsection*{Results -- SDC and crash cross-sections of the SW-only single-threaded baremetal benchmark basic tests}
Table~\ref{table:ps-cross-section} shows the estimated SDC cross-sections of the single-threaded baremetal benchmarks. Each benchmark ran more than 67k times, resulting in 3 hours of irradiation time per benchmark. The total beam time and fluence for all benchmarks were 18 hours and 6.12E+10  $n/cm^2$, respectively. Please note that we discarded the overhead time 
%due to system resets after crashes (i.e., the PS was unresponsive), 
required to configure and initialise the MPSoC  and collect the results from the DDR memory. 
%for the calculation of the SDC cross-section in Table~\ref{table:ps-cross-section}.

As expected, all benchmarks with a small memory footprint have either zero (see FFT, BasicMath, MatrixMul) or very low (see SHA) dynamic cross-sections. In contrast, the benchmarks with a large memory footprint (see QSort, CRC32) have the highest cross-section. We observe that Qsort is more vulnerable to SDCs than CRC32 despite its lower data segment size. This can be explained by the higher residence time of its data in the L2 cache. The data segment of Qsort fits in the 1~MB L2 cache of the APU and thus is not updated frequently from the off-chip DDR memory during execution, as done in  the case of the CRC32 benchmark. In contrast to the results of  \cite{anderson_redw_2018}, we report on average one order of magnitude higher dynamic cross-section for the single-threaded baremetal benchmarks; we tested the MPSoC on a broader range of benchmarks than \cite{anderson_redw_2018}, which exercised the APU caches in a more exhaustive way, thus revealing more errors. As mentioned, the authors in \cite{anderson_redw_2018} did not observe any SDC or crash but assumed one single error when calculating the dynamic cross-sections of single-threaded baremetal benchmarks running on the APU. However, we did not observe any processor crash, i.e. our findings in regards to the crash dynamic cross-section of the APU are the same as in \cite{anderson_redw_2018}.

\begin{table}
\centering
\caption{CPU benchmarks -- SDC cross-sections}
\resizebox{0.49\textwidth}{!}{
\begin{tabular}{lccccc} 
\hline

\textbf{Benchmark} & \textbf{Execution}  & \textbf{Fluence}  & \textbf{Total} &\textbf{SDC} & \textbf{SDCs} \\
                   & \textbf{time (s)}	& \textbf{($n/cm^2$)} & \textbf{runs}  &     & \textbf{cross-section} \\
\hline
FFT      & 1,227.95	& 6.96E+09 & 67,509 & 0	 & - \\ 
SHA	   & 1,239.14	& 7.02E+09 & 67,787 & 2	 & 2.85E-10 \\
BasicMath & 1,266.74	& 7.18E+09 & 67,940 & 0	 & - \\
MatrixMul & 1,556.26	& 8.82E+09 & 69,406 & 0	 & - \\
Qsort	   & 1,237.92	& 7.01E+09 & 67,487 & 38	 & 5.42E-09 \\
CRC32	   & 4,269.89	& 2.42E+10 & 67,572 & 18	 & 7.44E-10 \\ \hline
\textbf{Total}     & 10,797.90	& 6.12E+10 & 407,701& 58	 & 9.48E-10 \\
\hline
\end{tabular}
}
\label{table:ps-cross-section}
%\vspace{-0.4cm}
\end{table}

\section{Complex Tests}
\label{sec:Complex Tests}
This section presents the experimental methodology and results of the complex tests. These tests include two SW-only multi-threaded applications and one HW-SW co-design executing a CNN model, all running on top of the Linux OS.

\subsubsection*{Experimental setup}
The setup of the complex tests is the same as for the basic tests (see \figurename~\ref{fig:basic and undervolting tests setup}). However, the target board is not modified but instead powered by its onboard voltage regulators. In other words, we used the \textit{out-of-the-box} board (see Sec.~\ref{subsec:Target-board}) for the complex tests.

\subsubsection*{Testing procedure}
The Control-PC runs an in-house developed software, namely the Experiment Control Software (ECS), to orchestrate the test procedure of the target benchmarks through TCP/IP Ethernet.

The ECS software coordinates the tests of the applications via a shared Network File System (NFS) folder as follows: 1) the ECS initially resets the board and waits for it to boot, 2) after a successful OS boot, a \texttt{bash} script running on the MPSoC, namely, the \texttt{run.sh}, executes the following sub-tasks: 3a) connects on the shared NFS folder located on the Control-PC, 3b) updates a \texttt{sync.log} file in the NFS folder to notify the ECS of a successful OS boot, 3c) executes an initial run of the target benchmark to warm-up the CPU caches, 3d) notifies the ECS software via the \texttt{sync.log} file that it is ready to start running the benchmark, 3e) enters an infinite loop where it continuously runs the benchmark and stores the results in the NFS folder to be checked by the ECS. The execution and result checking (i.e., by the ECS) of each benchmark is synchronised with the ECS via a shared \texttt{mutex.log} file stored in the NFS folder. The ECS resets the board when it detects: 1) a boot timeout, 2) a critical error (classifying  an error as critical depends on the benchmark characteristics, as shown in the next section), or 3) a result query timeout.
It is worth noting that for each benchmark execution, the \texttt{run.sh} script saves the Linux \texttt{dmesg.log} of the target board for post-analysis to identify system-level errors, such as L1 and L2 cache errors (see section~\ref{subsec:SEUs-in-the-MPSoC-caches}). 
 
\subsection{SW-only multi-threaded applications
running under Linux OS}
\label{subsec:SW-only multi-threaded applications
running under Linux OS}
\subsubsection*{Benchmark details}
We tested two SW-only multi-threaded applications, namely the LFRic \cite{LFRiC} and the SVO \cite{SVO}, both running on top of the 4.19 Linux kernel, which was configured and compiled with PetaLinux 2019.2.

The LFRic is a weather and climate model and one of the H2020 EuroEXA project (http://euroexa.eu) target applications being developed by the UK's Met Office and its partners~\cite{LFRiC}. Much of the LFRic model's runtime consists of compute-intensive operations suitable for acceleration using FPGAs. The LFRic weather and climate model is based on the GungHo dynamical core with its PSyclone software technology~\cite{ADAMS2019383}. In our experiments, we exploited an essential computation kernel among the entire LFRic code, the matrix-vector product, to assess the overall dependability (i.e., dynamic cross-section) of the MPSoC. Specifically, this kernel supports 40-bit double-precision floating-point matrix-vector multiplications with an 8~$\times$~6 matrix and contributes significantly to the execution time of the Helmholtz solver that is used to compute atmospheric pressure~\cite{LFRiC}.

The SVO (Semi-direct Monocular Visual Odometry) processes raw data captured from visual sensors (e.g., camera) and conducts a probabilistic state estimation~\cite{SVO}. In particular, in the probabilistic state estimation, the algorithm calculates the camera's pose (i.e., motion estimation) and maps it to the surrounding, unknown environment (i.e. mapping). Both operations, the motion estimation and mapping are executed in parallel. SVO is used in many applications such as  
robotics and automotive applications to implement algorithms involving tasks like ego-motion or pose estimation of objects~\cite{SVO}.
%Finally, SVO has been the basis of several commercial applications, including Parrot-SenseFly Albris drone and autonomous car navigation by ZurichEye, now Facebook-Oculus VR Zurich. 

%The LFRic and SVO applications run on top of a Linux OS generated with AMD PetaLinux 2019.2. %All Arm A53 cores of the PS are configured to performance mode with NEON support disabled.

\subsubsection*{Results -- Error cross-sections of the SW-only multi-threaded applications}
\label{subsubsection:SW-only multi-threaded applications}

\begin{table}
\centering
\caption{SW-only multi-threaded Linux-based benchmark results}
\begin{tabular}{lcc} 
\hline
\textbf{Benchmark}            & LFRic       & SVO \\
\hline
Total runs         & 509       & 1,784 \\
Exec. time (hours)     & 4.3       & 6.5 \\ 
Soft-persistent crashes  & 6         & 39 \\
Recoverable crashes    & 20        & 94 \\
Total crashes          & 26        & 133 \\
Tolerable SDCs         &  0        & 51 \\
Critical SDCs          & 2         & 0 \\
Total SDCs             & 2         & 51 \\
Fluence (n/$\text{cm}^2$)	    &  9.35E+10 & 1.29E+11 \\
Total crash cross-section    & 2.78E-10  & 1.03E-09 \\
Total SDC cross-section      & 2.14E-11  & 3.96E-10 \\
\hline
\end{tabular}
\label{table:hpc-cross-section}
%\vspace{-0.55cm}
\end{table}

Table~\ref{table:hpc-cross-section} summarises the experimental results of the SW-only multi-threaded Linux-based benchmarks, which were collected during an 11-hour beam session. 

We categorise radiation-induced errors as crashes and SDCs. Crashes are further classified into \textit{soft-persistent} and \textit{recoverable} errors. Soft-persistent errors require several resets or a device power cycle to bring the MPSoC to a functional state. Recoverable errors require only one device reset to regain functionality. Similarly, SDC errors are classified into critical and tolerable as done in \cite{libano_tns_2019}. 
Critical errors lead to a result out of application specifications. Tolerable errors do not affect the final application result. 

Opposite to \cite{anderson_redw_2018}, which did not identify any SDC or processor hang (i.e., crash) when the APU was running multithreaded Linux-based benchmarks, our results showed that the MPSoC can experience radiation-induced errors. In detail, 5.11\% and 7.46\% of the total runs resulted in a crash for LFRic and SVO, respectively. 
From the total crashes of LFRiC, 23\% were soft-persistent, and 77\% were recoverable. For SVO, 29\% were soft-persistent and the remaining recoverable.
%We believe that the most probable reason that should have caused these soft-persistent crashes is an intermittent fault on the board's SD card interface. Nevertheless, we were not able to reproduce the intermittent fault after the radiation tests in our lab.
%An intermittent fault on the board's SD card interface must have caused the soft-persistent crashes.  

\begin{figure}[t]
\centerline{\includegraphics[width=3.2in]{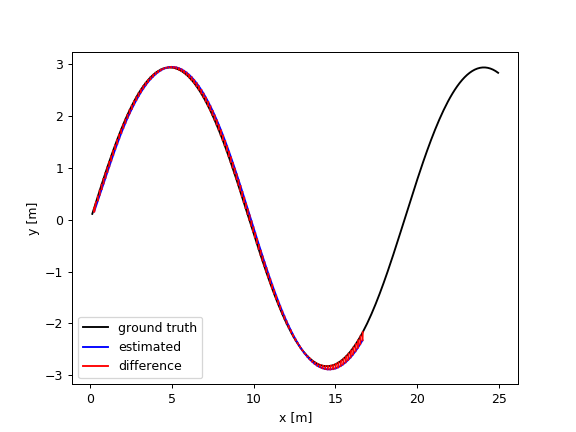}}
\caption{2D representation of the absolute trajectory error of an SVO run.}
\label{fig:error}
\vspace{-0.55cm}
\end{figure}

Regarding SDC errors, 0.39\% and 2.86\% of the total LFRic and SVO runs resulted in SDCs, respectively. 
However, our findings show that all SDCs of the SVO were tolerable and did not affect the correctness of the final application result. 
%The results mentioned above highlight the high severity that neutron particle strikes could have on the correct operation of the complex application. 
This can be justified by the inherent error resilience nature of computer vision algorithms like SVO, which commonly tolerate most SDCs. In other words, most SDCs cause a small deviation from the ground truth and, therefore, can be ignored. \figurename~\ref{fig:error} shows the absolute trajectory error of an SVO run under a tolerable SDC error. Although the result (i.e., estimated trajectory) deviated from the ground truth, it did not impact the in-field operation of SVO.
On the contrary, all SDCs for the LFRic application affected its final result and therefore were classified as critical. Commonly, the algorithmic nature of LFRic cannot tolerate any SDC.

% \begin{figure}[t]
% \centerline{\includegraphics[width=3.3in]{figures/jpeg/correct.png}}
% \caption{2D representation of the absolute trajectory error. Golden run.}
% \label{fig:correct}
% \end{figure}

\subsection{SW/HW multi-threaded co-design application
running under Linux OS}
\label{subsec:CNN-Accelerator}
This section includes results for the SW/HW co-design DPU from our previous study~\cite{dpu_radecs21}. We extend the study by providing the dynamic cross-section of crashes (i.e., hung) as well as the MTTF (see section~\ref{sec:Accessing the reliability of the MPSoC}) of the DPU application for different environments and device deployments. 

\begin{figure}[t]
\centerline{\includegraphics[width=3.2in]{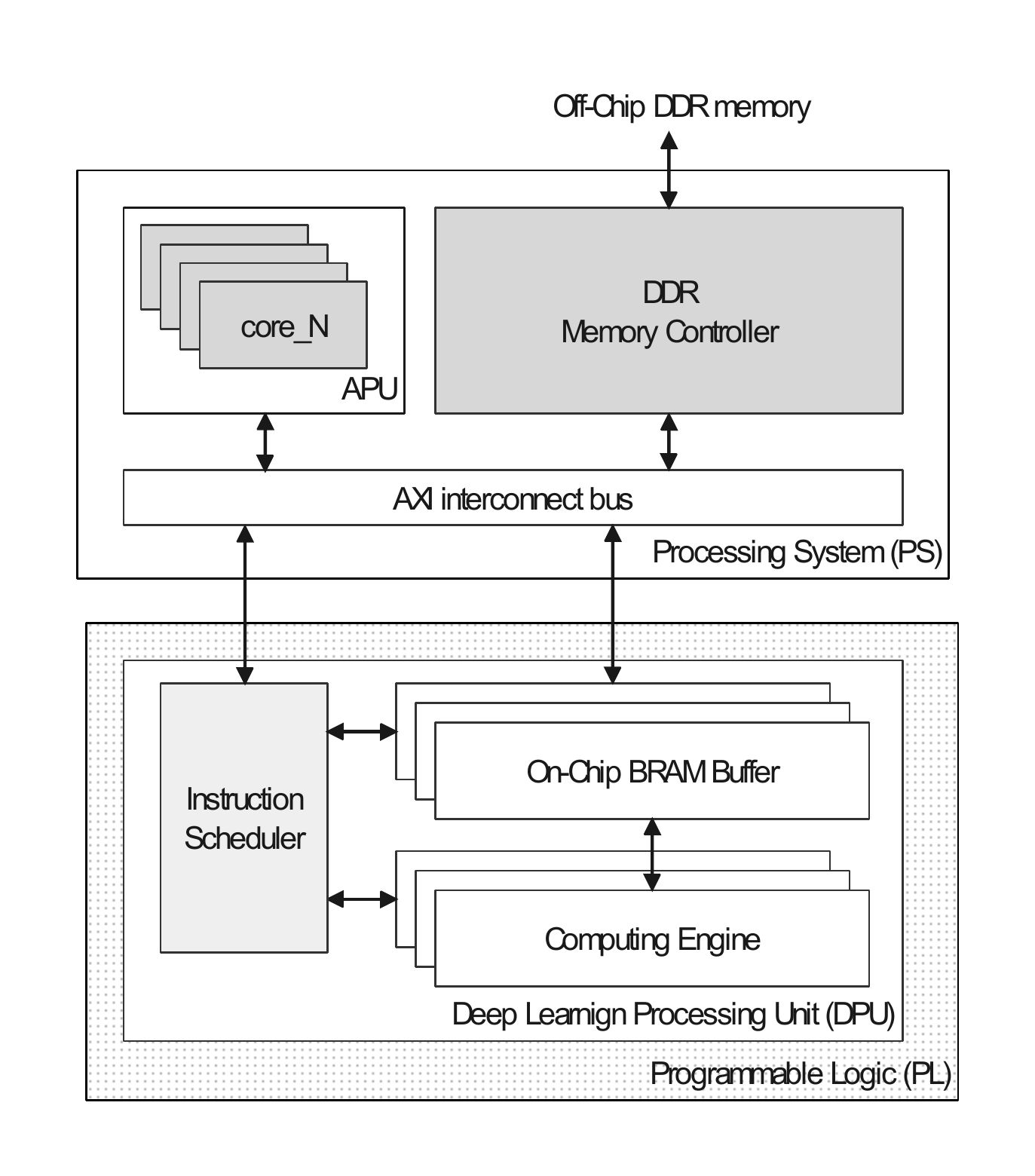}}
\caption{Deep-learing acceleration with the AMD Deep Processing Unit (DPU) on Zynq\textsuperscript{\textcopyright}-7000 SoC and Zynq\textsuperscript{\textcopyright} UltraScale+\textsuperscript{TM} MPSoC devices.}
\label{fig:DPU}
\end{figure}

\subsubsection*{Benchmark details}
\label{subsec:DPU}
We tested the MPSoC when running the \texttt{resnet50} image classification CNN model on the SW/HW Vitis AI DPU co-design. AMD has introduced a rich ecosystem of tools and IP accelerator cores to ease the development of AI applications. In more detail, AMD provides the Vitis AI development environment that encompasses 1) AI frameworks (e.g., Tensorflow), 2) pre-optimised AI models, 3) quantisation and model compression tools, and 4) the DPU with all necessary Linux drivers to seamlessly deploy a CNN application on AMD Zynq-7000 SoC and Zynq UltraScale+ MPSoC devices~\cite{UG338}.

The DPU accelerator is implemented with PL and is tightly interconnected via AXI interfaces to the PS, as shown in \figurename~\ref{fig:DPU}. The DPU executes special instructions that are generated by the Vitis AI compiler.
A typical Vitis AI development flow involves 1) the optimisation and compilation of a CNN model to DPU instructions and 2) the compilation of software running on the APU. 
The APU pre- and post-processes DNN data controls the DPU and orchestrates the movement of instructions and data between the DPU, the APU and the off-chip DDR memory.

The DPU consists of an instruction scheduler and up to three on-chip BRAM buffers and computing engines. The instruction scheduler fetches and decodes DPU instructions from off-chip memory and controls the on-chip memories and computing engines. The DPU is available in eight architecture configurations, i.e., B512, B800, B1024, B1152, B1600, B2304, B3136, and B4096. Each configuration utilises a different number of computing engines and on-chip memories to target different-size devices and support various DPU functionalities, e.g., ReLU, RELU6, or Leaky-ReLU.  

We implemented the Vivado DPU targeted reference design (TRD) \cite{vitis-1.3.1} provided by Vitis AI v1.3.1 with Vivado 2020.2 for our target board (i.e., ZCU102). The DPU was synthesised with default settings, i.e., \texttt{B4096} convolution architecture with \texttt{RAM\char`_USAGE\char`_LOW, CHANNEL\char`_AUGMENTATION\char`_ENABLE, DWCV\char`_ENABLE, POOL\char`_AVG\char`_ENABLE, RELU\char`_LEAKYRELU\char`_RELU6}, and \texttt{Softmax}. The parallelism of the DPU can be defined in three dimensions, input channel parallelism (ICP), output channel parallelism (OCP), and pixel parallelism (PP).  
The \texttt{B4096} architecture has ICP and OCP equal to 16, PP equal to 8, and can achieve up to 4096 operations per clock cycle. The \texttt{RAM\char`_USAGE\char`_LOW} configuration utilises 257 BRAM36 primitives for buffering weights, bias and intermediate features. Channel augmentation improves the DPU utilisation when the number of input channels is much lower than the available channel parallelism. DepthwiseConv (DWCV), AveragePool, LEAKYRELU and RELU6 are standard CNN parameters that are described in \cite{UG338}.
The design was implemented with Vivado's \texttt{Performance\char`_ExplorePostRoutePhysOpt} run strategy because Vivado's \texttt{default} run strategy resulted in time violations for the default operating frequencies of the implemented TRD.
Table~\ref{table:resource-util} shows the resource utilisation and operating frequency of the DPU TRD. Vivado reported that 41.45\% (i.e., 59,281,993 bits) of the device's configuration bits
were essential. Please recall that \emph{essential bits} are configuration bits that, when corrupted, can potentially cause functional errors in the application.

\begin{table}
\centering
\caption{Resource utilisation and operating frequency of the DPU SW/HW co-design application}
%\setlength{\tabcolsep}{5pt}
\begin{tabular}{lrrrr}
\hline
Resource & Utilisation & Available & Utilisation  & Frequency \\
\hline 
LUT   & 108,208 & 274,080 & 39.48 \% & 325 MHz \\
LUTRAM & 11,960 & 144,000 & 8.31 \%  & 325 MHz \\
FF    & 203,901 & 548,160 & 37.20 \% & 325 MHz \\
BRAM  & 522   & 912   & 57.24 \% & 325 MHz \\
DSP   & 1,395 & 2,520 & 55.36 \% & 650 MHz \\
IO    & 7     & 328   & 2.13 \%  & 325 MHz \\
BUFG  & 6     & 404   & 1.49 \%  & 325 MHz \\
MMCM  & 1     & 4     & 25.00 \% & 325 MHz \\
PLL   & 1     & 8     & 12.50 \% & 325 MHz \\
APU   & 1     & 1     & 100.00 \% & 1200 MHz \\
DDR ctrl. & 1     & 1     & 100.00 \% & 533  MHz \\
\hline
\end{tabular}%
\label{table:resource-util}
%\vspace{-0.6cm}
\end{table}

Two important notes can be made for Table~\ref{table:resource-util}. 
First, all resources in the DPU operate at 325~MHz except for the DSPs, which run at 2~$\times$~325~MHz~=~650~MHz. This is because the DPU design applies a double data rate technique on DSP resources. Since DSPs can operate at a much higher frequency than other PL resources, one can perform \emph{N} times more computation by running the DSPs with \emph{N} times the frequency of the surrounding logic while multiplexing and de-multiplexing their input and output data, respectively. 

Second, the design utilises \texttt{319, 55, 405, 4} and \texttt{1} LUT, LUTRAM, FF, BRAM and DSP more primitives than the baseline TRD design. This is because we included the AMD SEM IP in the design to perform fault injection and validate our experimental setup before the radiation experiments. 
However, we turned scrubbing off (configured SEM IP to IDLE mode) during beamtime to allow the DPU to accumulate at least one CRAM upset per image classification.   Otherwise, the DPU would have performed almost all classifications without a CRAM upset. The SEM IP operating at 200MHz would have recovered  much faster CRAM upsets (1700 upsets per minute) than they occurred (8 upsets per minute -- estimated for the 5.6E+6~neutrons/cm\textsuperscript{2}/seconds neutron flux at ChipIR facilities).
Instead of scrubbing the device, all CRAM upsets recovered after a device reset when the DPU reported a tolerable or non-tolerable error or a crash (i.e., timeout).
%Each image classification lasted 1.7 seconds, from which 10~mS were devoted to the actual DPU execution and the remaining to initialise the DPU and move data between the DPU and PS.  

We used Petalinux 2020.2 to generate a Linux OS image for the ZCU102 by using the default Board Support Package (BSP) provided by the \texttt{DPU-TRD}, except 1) the \texttt{nfs\_utils} package, which was additionally enabled to mount an NFS folder on Linux, and 2) the u-boot bootloader configuration which mounted an \texttt{EXT4} file system on an SD card instead of an \texttt{INITRD} RAM disk on the DDR memory.

The CNN application that ran on the DPU was the 8-bit quantised, not pruned \texttt{resnet50.xmodel}, provided by the Vitis AI TRD. 
%but serves nice as a baseline application for comparison with more optimised models that we aim to implement and test in future work.

\subsubsection*{Results -- Neutron error (SDC and crash) cross-sections of AMD Vitis DPU running image classification}
%In this section, we discuss the impact of neutron radiation effects on the reliability of the DPU accelerator. Given that the DPU comprises of a heterogeneous architecture including the ARM SoC and the FPGA fabric, we first present the cross-sections of the memories of the PS part (i.e. CPU caches) and then discuss how the  SEUs in the PL configuration memory affect the behaviour of the system. Please note that the neutron radiation experiments took place at ChipIr on May 2021.
Table~\ref{tab:dpu-cross-section} shows the dynamic cross-section of the DPU running the resnet50 image classification CNN for a total fluence of \texttt{5.5x10\textsuperscript{10}} neutrons/cm\textsuperscript{2} during a 3-hour radiation test session.
The DPU accelerator performed 5985 classification runs in total, from which 50\% of the runs resulted in an SDC, 1.5\% in a crash, and 49.5\% were correct. Only 1.57\% of the total SDCs resulted in image misclassification or, in other words, were critical. The experimental results show a reliable operation of the DPU even though it did not incorporate any soft error masking scheme in its PL logic like triple modular redundancy (TMR)~\cite{NMR_Rel} or ECC in its utilised BRAMs~\cite{ECC_Rel}. 

However, the dynamic cross-section of the DPU is not only affected by soft errors in its PL part but also due to errors in the APU. 
As mentioned, the DPU is an SW/HW co-design, which means that both the APU and PL logic should cooperate in a reliable manner to successfully classify an image when running the resnet50 model. In the following, we measure the effectiveness of all soft-error mitigation schemes embedded in the APU to cope with upsets in the L1 and L2 caches of the processor. 
%demonstrating that the resnet50 application can reliably run on the DPU even when upsets are accumulated in the MPSoC.
%WE NEED TO SAY ABOUT RESETS. HOW MANY CRACHES WE had.
%Notice that our SW/HW complex tests corrected upsets by resetting the board every 20 minutes. Thus, memory upsets accumulate in the MPSoC until the next reset cycle of the system. These accumulated upsets may cause the system to crash or produce SDCs in consecutive runs. We removed all these abnormal results due to accumulated errors from the SDC cross-section calculation of the resnet50 application. 
%We used the CRAM and BRAM cross-sections reported in Section~\ref{sec:BasicTests} and the utilised programmable resources of the DPU (i.e., essential configuration bits and BRAMs used) to estimate how many upsets accumulated per DPU classification. These are 0.10 and 0.22 upsets per classification in the essential CRAM bits and BRAM contents of the MPSoC, respectively. Therefore, each classification run that lasted 1.7 seconds experienced, on average, 0.32 upsets. Additionally, we estimated that more than 6 upsets were accumulated during the 38 seconds OS boot and warm-up period of each reset cycle. Thus, the DPU circuit encountered more than one upset in the PL memories for all classification runs. Consequently, we consider the ratio of critical errors (0.78\%) as an upper bound of the DPU's AVF.

\begin{table}[htbp]
  \centering
  \caption{Neutron SDC cross-section of AMD Vitis DPU running image classification}
    \begin{tabular}{|l|r|r|r|r|r|}
          \hline
          & \multicolumn{2}{c|}{Classification}  & \multicolumn{1}{c|}{Cross} & \multicolumn{2}{c|}{Conf. Level} \\
          & \multicolumn{2}{c|}{runs}  & \multicolumn{1}{c|}{Section} & \multicolumn{2}{c|}{95\%} \\
          & \multicolumn{1}{c|}{\#} & \multicolumn{1}{c|}{\%} 
          & \multicolumn{1}{c|}{(cm\textsuperscript{2})} & \multicolumn{1}{c|}{Lower} & \multicolumn{1}{c|}{Upper} \\
          \hline \hline
    Correct runs & 2964  & 49.52\% & \multicolumn{1}{c|}{-} & \multicolumn{1}{c|}{-} & \multicolumn{1}{c|}{-} \\
    Crashes & 89  & 1.49\% & 1.60E-09
 & 1.26E-09
 & 2.02E-09 \\
    Critical (C)  & 46    & 0.77\% & 8.29E-10 & 6.07E-10 & 1.11E-09 \\
    Tolerable (T) & 2886  & 48.22\% & 5.20E-08 & 5.01E-08 & 5.39E-08 \\
    C+T errors  & 2932  & 49.99\% & 5.28E-08 & 5.09E-08 & 5.48E-08 \\ \hline
    \end{tabular}%
  \label{tab:dpu-cross-section}%
\end{table}

\subsubsection*{Results -- MPSoC APU L1 and L2 cache cross-section when running image classification with the AMD Vitis DPU}
\label{subsec:SEUs-in-the-MPSoC-caches}
We post-processed the Linux \texttt{dmesg.log} files captured during the AMD DPU tests to analyse the NSEUs observed in the MPSoC APU caches. We report the cross-sections of Level-1 Data (L1-D) and Instruction (L1-I) caches, Translation Lookaside Buffer (TLB), Snoop Control Unit (SCU), and Level-2 cache. Moreover, the upsets in the data and tag arrays in both the L1 and L2 caches have been separately identified.

In detail, Table~\ref{tab:l1c-cross-section} shows the dynamic cross-sections of the 32~KB L1-D cache, the 32~KB L1-I cache, and the TLB -- a two-level TLB with 512 entries that handles all translation table operations of the APU. 

\begin{table}[htbp]
  \centering
  \caption{L1 Cache Cross-Section}
    \begin{tabular}{|l|r|r|r|r|r|}
          \hline
          & \multicolumn{1}{c|}{Size} 
          & \multicolumn{1}{c|}{Upsets} & \multicolumn{1}{c|}{Cross-sec.} & \multicolumn{2}{c|}{Conf. Level 95\%} \\
          &\multicolumn{1}{c|}{(bit)}
          & \multicolumn{1}{c|}{(bit)}
          & \multicolumn{1}{c|}{(cm\textsuperscript{2}/bit)} & \multicolumn{1}{c|}{Lower} & \multicolumn{1}{c|}{Upper} \\
          \hline \hline
    L1-D Data & 262,144& 32    & 2.20E-15 & 1.50E-15 & 3.11E-15 \\ 
    L1-D Tag & 155,648& 3     & 3.47E-16 & 7.16E-17 & 1.02E-15 \\ 
    L1-D Total & 417,792& 35    & 1.51E-15 & 1.05E-15 & 2.10E-15 \\ 
    L1-I Data & 262,144& 25    & 1.72E-15 & 1.11E-15 & 2.54E-15 \\ 
    L1-I Tag & 147,456& 4     & 4.89E-16 & 1.33E-16 & 1.25E-15 \\ 
    L1-I Total & 409,600& 29    & 1.28E-15 & 8.54E-16 & 1.83E-15 \\ 
    L1 TLB & 16,384& 9     & 9.90E-15 & 4.53E-15 & 1.88E-14 \\ \hline
    \end{tabular}%
  \label{tab:l1c-cross-section}%
  %\vspace{-0.1cm}
\end{table}%

\begin{table}[htbp]
  \centering
  \caption{L2 Cache Cross-Section}
    \begin{tabular}{|l|r|r|r|r|r|}
          \hline
          & \multicolumn{1}{c|}{Size} 
          & \multicolumn{1}{c|}{Upsets} & \multicolumn{1}{c|}{Cross-sec.} & \multicolumn{2}{c|}{Conf. Level 95\%} \\
          &\multicolumn{1}{c|}{(bit)}
          & \multicolumn{1}{c|}{(bit)} & \multicolumn{1}{c|}{(cm\textsuperscript{2}/bit)} & \multicolumn{1}{c|}{Lower} & \multicolumn{1}{c|}{Upper} \\
          \hline \hline
    L2 Data & 8,388,608 &293   & 6.29E-16 & 5.59E-16 & 7.06E-16 \\
    L2 Tag & 4,194,304 &20    & 8.59E-17 & 5.25E-17 & 1.33E-16 \\
    L2 Total & 12,582,912 &313   & 4.48E-16 & 4.00E-16 & 5.01E-16 \\
    SCU & 155,648 &4     & 4.63E-16 & 1.26E-16 & 1.19E-15
    \\ \hline
    %\multicolumn{5}{l}{*The cross-section for SCU is given in cm\textsuperscript{2} per device because} \\
    %\multicolumn{5}{l}{we do not know the memory size of SCU.}
    \end{tabular}%
  \label{tab:l2c-cross-section}%
  %\vspace{-0.4cm}
\end{table}%

Table~\ref{tab:l2c-cross-section} presents the cross-sections of the 1~MB Level-2 cache (L2)  and the SCU. The SCU has duplicate copies of the L1 data-cache tags. It connects the APU cores with the device's accelerator coherency port (ACP) to enable hardware accelerators in the PL to issue coherent accesses to the L1 memory space.
The cross-sections of the tag arrays have been calculated based on the tag sizes of the caches, e.g., a 16-bit tag in the 16-way set associative, 64-byte line, 1~MB L2 cache. 
As mentioned, the cross sections have been calculated for a total fluence of \texttt{5.55x10\textsuperscript{10}} neutrons/cm\textsuperscript{2}.
The results show that the cross-sections of the tag arrays are slightly lower than those of the data arrays. The average cross-section calculations for all caches (i.e., L1 and L2) in the MPSoC are close to those reported by Jordan D. Anderson et al. in \cite{anderson_redw_2018}. 

\begin{figure}[t]
\centerline{\includegraphics[width=3.0in]{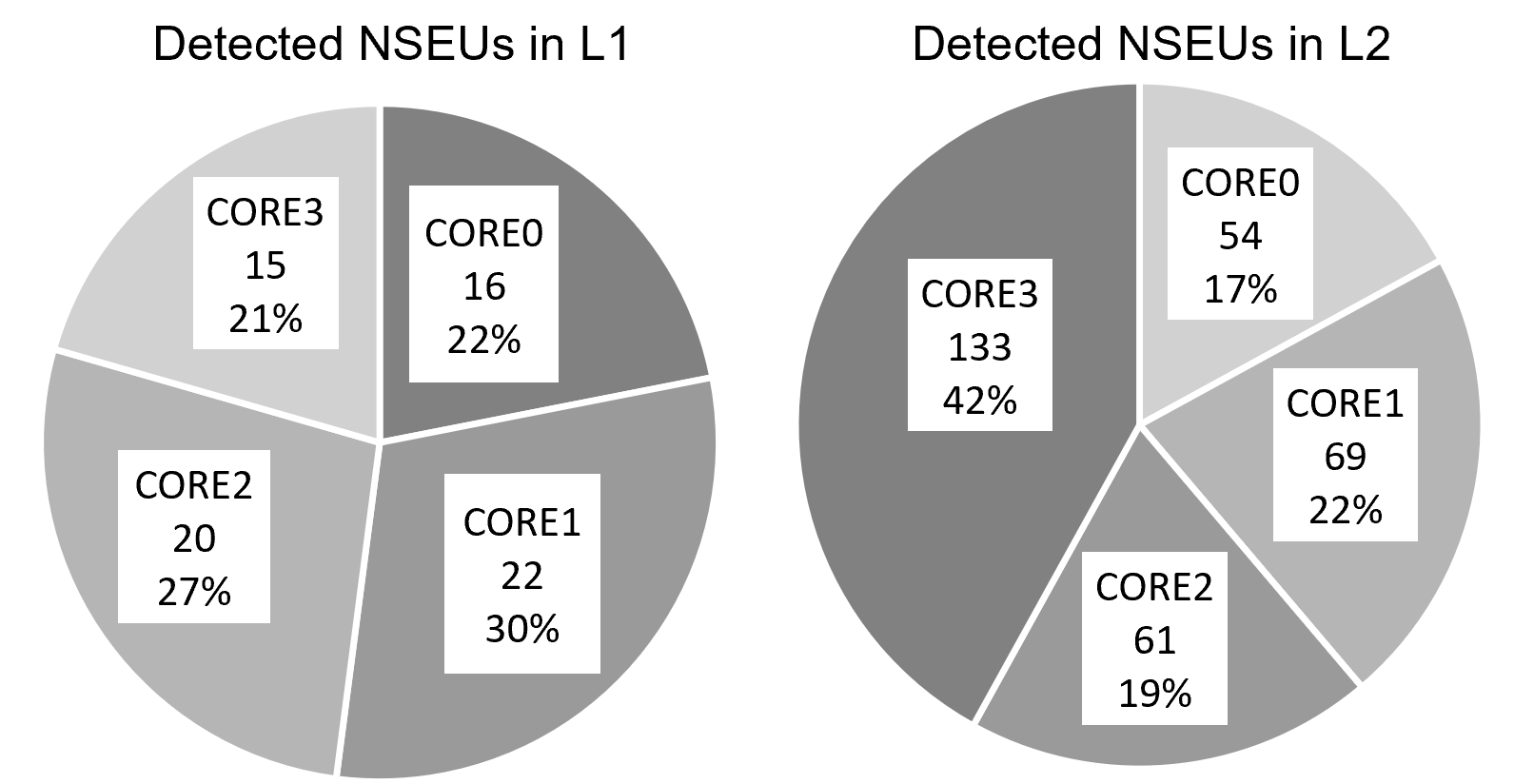}}
\caption{Detected cache upsets per APU Core.}
\label{fig:cache-upsets-per-core}
\vspace{-0.65cm}
\end{figure}

\figurename~\ref{fig:cache-upsets-per-core} presents the number of detected upsets per cache per APU core. The upsets in the L1 caches are balanced between the four cores, while in the L2 cache, more upsets were observed in the 3\textsuperscript{rd} APU core of the MPSoC. We assume that the Linux OS utilised more Core-3, and thus more cache upsets were detected for Core-3 in the L2 cache.

The private L1-I caches are protected against NSEUs with parity checking (i.e., only error detection is supported), while the private L1-D caches and the shared L2 cache feature SECDED via ECC. However, we observed crashes and SDCs during image classifications with the DPU (and also in the SW-only basic and complex tests) despite the soft error mitigation mechanisms incorporated in the APU caches. We reason that the application errors occurred due to uncorrectable errors in the APU caches (e.g., double-bit errors within a memory word slice of the L1 or L2 caches protected by the same parity bits) or due to upsets in the configuration bits of the PL in case of the DPU. For example, SBUs in L1-D and L2 caches are successfully detected and corrected through SECDED mechanisms, while SBUs in L1-I caches are detected through parity checking and repaired by invalidating and reloading the cache. Similarly, double-bit upsets in L2 are detected by the SECDED scheme and corrected  with cache invalidation to force a cache update from a lower memory hierarchy, e.g., DDR.   However, if a double-bit error affects a ``dirty'' line of a write-back L1-D and L2 cache, its data is lost, resulting in data corruption. In case of double-bit upsets in the parity-protected L1-I caches, these cannot be detected.

\section{Accessing the reliability of the MPSoC}
\label{sec:Accessing the reliability of the MPSoC}
In sections~\ref{sec:BasicTests}\&\ref{sec:Complex Tests}, we calculated the static and dynamic cross sections of the XCZU9EG in various scenarios under neutron accelerated radiation testing, e.g., when executing a simple SW-only baremetal single-threaded benchmark or complex Linux-based SW/HW co-design application for image classification.  In this section, we project the measured cross-sections of the XCZU9EG at different terrestrial radiation environments and device deployments and examine the reliability of the MPSoC-based computing system under the lens of the MTTU and MTTF dependability metrics as described in section~\ref{subsec: Cross-section and Failure Rate of Semiconductor Devices}.

\begin{figure}
\centering
    \begin{subfigure}[b]{0.50\textwidth}
    %\centerline{\includegraphics[trim=0cm 3cm 0cm 4cm, width=4in]
    %trim=left botm right top
    \centerline{\includegraphics[trim=0cm 0cm 0cm 0cm, width=0.95\textwidth]{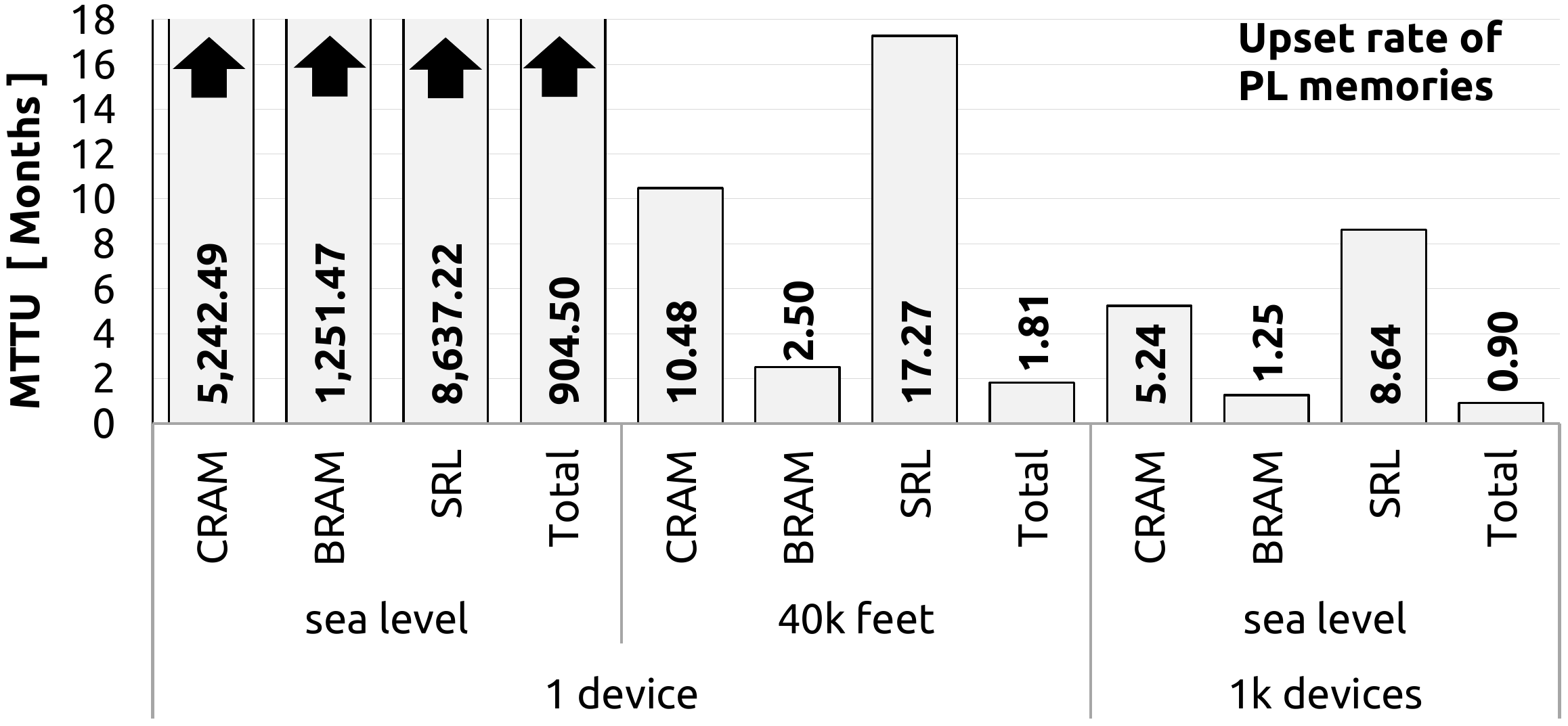}}
	\vspace*{0mm}
    \caption{} \label{fig:MTTU-PL-Memory}
\end{subfigure}
\begin{subfigure}[b]{0.50\textwidth}
    %trim=1cm 4cm 0.0cm 4.0cm, 
    \centerline{\includegraphics[trim=0.3cm 0cm 0cm 0cm, width=0.95\textwidth]{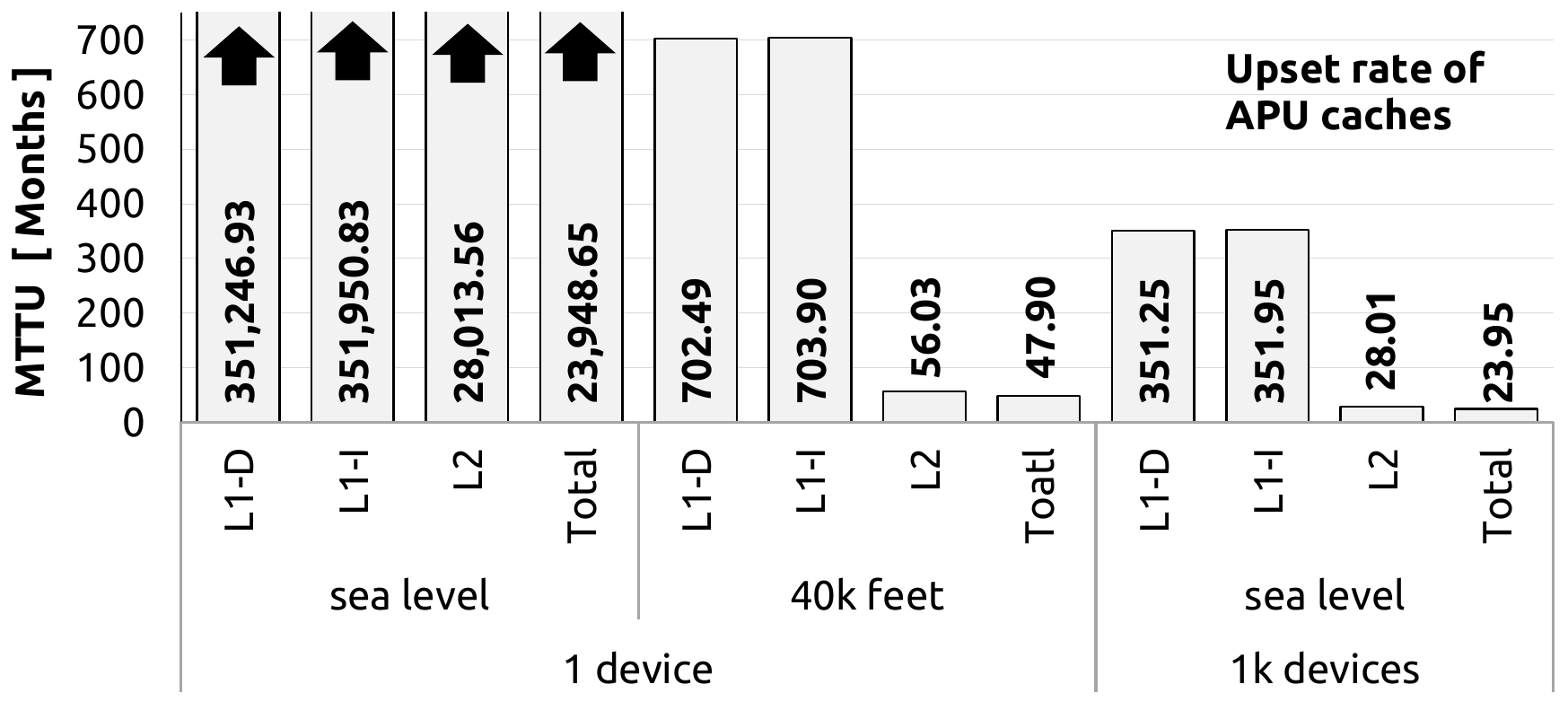}}
	\vspace*{-2mm}
    \caption{} \label{fig:MTTU-APU-ALL-Caches}
\end{subfigure}\hfill
\caption{(a) MTTU in PL memories measured for the simplex tests, (b) MTTU of the APU L1 data (L1-D), L1 instruction (L1-I) and L2 caches when running the DPU SW/HW co-design. The MTTU metrics have been calculated for a system with one MPSoC operating in NYC at sea level or 40k altitude and a system using 1000 MPSoCs in NYC at sea level.}
\label{fig:MTTU-PL-APU-CACHES}
\vspace{-0.2cm}
\end{figure}

\figurename~\ref{fig:MTTU-PL-APU-CACHES}~(a) shows the MTTU of the MPSoC's PL memories assuming 1) a computing system that uses one MPSoC and operates at NYC sea level (e.g., an automotive application), 2) at 40k feet altitude (e.g., avionics), and 3) a system that uses 1k MPSoC devices and operates at the NYC sea level (e.g., a 1000 MPSoC node data centre). 
%Please note that the MTTU and MTTF results for operation at the sea level are calculated by dividing the MTTF figures by 500. Similarly, the MTTF for the 1000-node MPSoC system is computed by dividing the MTTF figures by 1000.

On average, the system consisting of one MPSoC and operating at sea level will experience a neutron-induced upset in the CRAM, BRAM or SRL memories of the device every 904 months (i.e., 75 years). 
%, assuming that it utilises all the available programmable resources. 
However, the MTTU (i.e. upset rate) of the PL memories of the same system operating at 40k feet altitude drops to 1.81 months (i.e., 500\texttt{X} reduction). On the other hand, a system consisting of 1k MPSoC computing nodes will collectively encounter one upset in PL memories every 0.9 months on average. %Additionally, we calculate that the MTTU of PL memories deteriorates by 26.4\% and 35.6\% when the supply voltage drops to 0.75V and 0.60V, respectively, in any of the above systems.
The MTTU results show that fault-tolerance techniques such as configuration memory scrubbing and ECC in BRAMs should be considered in MPSoC systems that operate at high altitudes or on a large scale (i.e., data centres) to avoid the accumulation of upsets in its PL memories. %Moreover, engineers should take into account the negative effects of supply voltage scaling since the increased upset rate may negate the power saving of undervolting techniques in MPSoCs.  

\figurename~\ref{fig:MTTU-PL-APU-CACHES}~(b) illustrates the MTTU of the L1-D, L1-I and L2 caches of the MPSoC's APU when running the SW/HW DPU co-design. In other words, the cache upset rates of the APU were calculated by using the dynamic cross-section of caches in the DPU application. As expected, the MTTU of the APU caches is 26.5\texttt{x} higher than the PL memories due to their much smaller size. We calculated that the MTTU of caches in the one- and 1k-node(s) system could drop to 48 and 24 months, respectively, which points out that the parity and SECDED mechanisms of the APU are a necessary feature in the MPSoC, especially when used in large scale systems. The effectiveness of these embedded soft-error mitigation mechanisms is evaluated in the following sections, where we measure the dynamic cross-section of various MPSoC applications, i.e., report the rate at which memory upsets could not be recovered,  thus resulting in an SDC or processor crash.

\begin{figure}
%trim=left botm right top
%\centerline{\includegraphics[trim=0.5cm 4cm 0.5cm 4.0cm, width=4in]
\centerline{\includegraphics[width=0.5\textwidth]{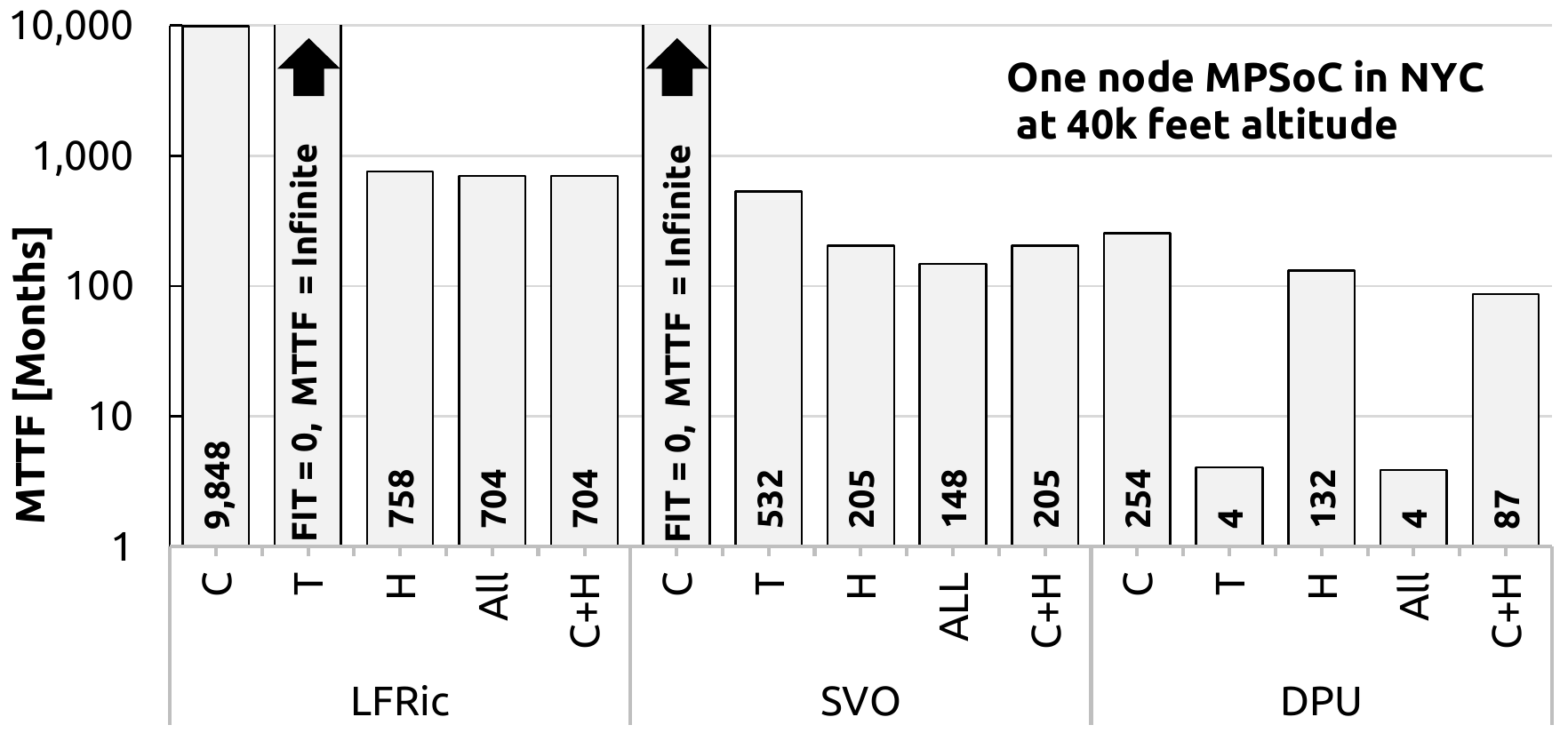}}
%\vspace{-2.0cm}
\caption{MTTF of 1) the SW-only multi-threaded applications (LFRic, SVO), and 2) the SW/HW multi-threaded co-design application (DPU). The MTTF metrics have been for one MPSoC-based computing system operating in NYC at 40k feet.}
\label{fig:MTTF-APPS}
\vspace{-0.5cm}
\end{figure}

Our analysis shows that the MPSoC has a low upset rate in PL memories and even lower in APU caches when operating in a single node computing system in NYC at sea level and increases in systems operating at high altitudes or on a large scale. In the following, we present the MTTF of MPSoC applications operating in a relatively high neutron flux to understand how an increased upset rate can affect reliability at the application level.
In detail, \figurename~\ref{fig:MTTF-APPS} presents the MTTF of the MPSoC when running the SW-only multi-threaded applications (i.e., LFRic and SVO) and the SW/HW DPU co-design. The MTTF of all applications is calculated assuming operation in NYC at 40k feet altitude.  However, the MTTF figures for operation at the sea level or for the 1000-node MPSoC system can be calculated by dividing and multiplying the MTTF figures of \figurename~\ref{fig:MTTF-APPS} by 500, respectively. 

As mentioned in section~\ref{sec:Complex Tests}, errors of the complex tests have been categorised into critical SDCs (C), tolerable SDCs (T), and processor hang (H) or otherwise crash. An application failure occurs during an SDC or a processor hang event. In this case, the overall FIT rate of the system is
\begin{equation}
   \text{FIT}_{\text{all}} = \text{FIT}_{\text{critical}} + \text{FIT}_{\text{tolerable}} + \text{FIT}_{\text{hang}}
\end{equation}
However, in error-resilient applications, we can omit the $\text{FIT}_{\text{tolerable}}$ from our calculations since tolerable SDCs do not affect output correctness. Thus, the overalls FIT can be calculated as follows:
\begin{equation}
   \text{FIT}_{\text{C+H}} = \text{FIT}_{\text{critical}} + \text{FIT}_{\text{hang}}
\end{equation}
In \figurename~\ref{fig:MTTF-APPS} the MTTF of $\text{FIT}_{\text{all}}$ is refered as \texttt{All} and for $\text{FIT}_{\text{C+H}}$ as \texttt{C+H}. 

Regarding the MTTF results, we see that the failure rate of the SW-only LFRiC and SVO applications is, on average, one order of magnitude lower than the rate of upsets in APU L2 caches. This shows that the embedded SECDED mechanisms in the APU are effective even for a high upset rate in caches. Although the upset rate in the caches has been calculated for the DPU SW/HW co-design, we believe similar figures would hold for the LFRiC and SVO applications. All complex tests share the same operating system and use the same software to send and receive data from the control PC. Therefore we expect that the caches would be exercised similarly in all benchmarks and thus have the same dynamic cross-section. However, the MTTF\textsubscript{All} of SVO is 79\% lower than LFRiC, because SVO is more vulnerable to cache upsets due to its larger memory footprint. On the other hand, as mentioned in section~\ref{subsec:SW-only multi-threaded applications running under Linux OS}, all SDCs in LFRic are critical, while in SVO tolerable. Thus, the reliability degradation of SVO w.r.t. to LFRiC can be limited to 70\% if we omit the FIT rate of tolerable SDCs from SVO, i.e. if we consider the MTTF\textsubscript{C+H} of the applications.  

Comparing the SW/HW co-design (i.e., DPU) with the SW-only applications (i.e., BareC, LFRic, and SVO), we observe that the DPU has, on average, 90\texttt{x} lower MTTF\textsubscript{All}. This can be justified due to the high FIT rate (low MTTF) of the PL accelerator, which deteriorates the total MTTF of the SW/HW co-design application. In contrast, BareC, LFRic, and SVO do not integrate any PL accelerator and therefore have an overall higher MTTF than the DPU.

However, the MTTF\textsubscript{All} of the DPU is very low due to the increased rate of tolerable SDCs. Omitting the FIT rate of tolerable SDCs yields an MTTF\textsubscript{C+H} = 87 months, which is 4\texttt{x} lower than the MTTF\textsubscript{C+H} of the SW-only applications. The MTTF results of the DPU show that deploying SW/HW co-design applications at high altitudes or on a large scale requires some form of soft error mitigation like configuration memory scrubbing or even hardware redundancy in high-reliability systems.

\section{Conclusions}
\label{sec:Conclusions}
This article evaluated the neutron Single Event Effect (SEE) sensitivity of the AMD UltraScale+ XCZU9EG MPSoC through accelerated neutron radiation testing and dependability analysis. The cross sections of the device's Programmable Logic (PL) and Processing System (PS) memories were characterised under the following workloads: 1) a synthetic design that utilised all PL resources, 2) several single-threaded baremetal SW-only benchmarks, 3) two SW-only multi-threaded Linux-based applications for weather prediction and pose estimation, and 4) a SW/HW DPU co-design running the resnet50 image classification model. The device's neutron CRAM static cross-section was measured to be 1.84E-16cm\textsuperscript{2}/bit, which is in the range of previous studies (1.10E-16 cmcm\textsuperscript{2}/bit -- 3.40E-16 cmcm\textsuperscript{2}/bit). The cross-sections of BRAM and SRL memories were one order of magnitude higher than CRAM.   No NSEU in the CRAM resulted in a Multi-Cell Upset (i.e., two or more upsets in one configuration frame), concluding that SECDED scrubbing is adequate to recover PL upsets in XCZU9EG devices when used in terrestrial applications. We observed only one BRAM SEFI, one SRL SEFI and two SELs during the accelerated radiation tests, which exposed the MPSoC to more than 1.3 million hours of equivalent natural neutron fluence at NYC sea level. We conclude that the probability of SEFIs and SELs in MPSoC terrestrial applications is extremely low.

To put the cross-section measurements into context, we conducted a dependability analysis assuming a one-node MPSoC system operating at NYC sea level (e.g., automotive) or 40k altitude (e.g., avionics) and a 1000-node MPSoC system at NYC sea level. All SW-only benchmarks achieved a MTTF higher than 148 months in the one-node system at 40k altitude, which points out that the PS can operate reliably despite a relatively high rate of cache upsets (MTTU = 48 months). Thus, we conclude that the embedded SECDED mechanisms of the PS can effectively recover NSEUs even in high altitude or large-scale MPSoC systems. However, the DPU application was more prone to neutron-induced errors than the SW-only workloads. The MTTF of the DPU was estimated to be 4 months, assuming it runs on the same one-node system at sea level. Thus, we conclude that SW/HW applications require extra soft error mitigation, e.g., hardware redundancy, to improve reliability in particular environments and device deployments. Finally, we showed that error-resilient applications like the DPU image classification can ignore tolerable errors to improve MTTF since these do not affect the final system result. 

\bibliographystyle{IEEEtran}
\bibliography{citations}
\end{document}